\documentclass[aps,prapplied,twocolumn,superscriptaddress]{revtex4-2}

\usepackage{siunitx}
\usepackage{soul,color,xcolor}
\usepackage{mhchem}
\usepackage{enumerate}
\usepackage{graphics}
\usepackage{graphicx}
\usepackage{multirow}

\begin{document}

\title{Hierarchical Recording Architecture for Three-Dimensional Magnetic Recording}

\author{Yugen Jian}
\affiliation{\mbox{Wuhan National Laboratory for Optoelectronics, Huazhong University of Science and Technology, Wuhan 430074, China}}
\affiliation{\mbox{Department of Electrical and Computer Engineering, National University of Singapore, Singapore 117583, Singapore}}
\author{Ke Luo}
\affiliation{\mbox{Wuhan National Laboratory for Optoelectronics, Huazhong University of Science and Technology, Wuhan 430074, China}}
\author{Jincai Chen}
\email{jcchen@hust.edu.cn}
\affiliation{\mbox{Wuhan National Laboratory for Optoelectronics, Huazhong University of Science and Technology, Wuhan 430074, China}}
\affiliation{\mbox{School of Computer of Science and Technology, Huazhong University of Science and Technology, Wuhan 430074, China}}
\affiliation{\mbox{Key Laboratory of Information Storage System, Engineering Research Center of Data Storage Systems and Technology}, Ministry of Education of China, Wuhan 430074, China}
\author{Xuanyao Fong}
\email{kelvin.xy.fong@nus.edu.sg}
\affiliation{\mbox{Department of Electrical and Computer Engineering, National University of Singapore, Singapore 117583, Singapore}}

\begin{abstract}
Three-dimensional magnetic recording (3DMR) is a highly promising approach to achieving ultra-large data storage capacity in hard disk drives. One of the greatest challenges for 3DMR lies in performing sequential and correct writing of bits into the multi-layer recording medium. In this work, we have proposed a hierarchical recording architecture based on layered heat-assisted writing with a multi-head array. The feasibility of the architecture is validated in a dual-layer 3DMR system with \ce{FePt}-based thin films via micromagnetic simulation. Our results reveal the magnetization reversal mechanism of the grains, ultimately attaining appreciable switching probability and medium signal-to-noise ratio (SNR) for each layer. In particular, an optimal head-to-head distance is identified as the one that maximizes the medium SNR. Optimizing the system's noise resistance will improve the overall SNR and allow for a smaller optimal head-to-head distance, which can pave the way for scaling 3DMR to more recording layers.
\end{abstract}

\maketitle

\section{INTRODUCTION
\label{sec1}}
The advancement and productization in generative artificial intelligence (AI) models have become the catalyst for a data deluge these years, intensifying a huge demand for data storage devices \cite{2024(wright)worldwide}. Mass-capacity hard disk drives (HDDs) continue to be one sustainable and reliable choice to scale against resource scarcity in physical space, power consumption, and budget for enterprise data centers \cite{2024(burns)worldwide,2024(seagatetechnology)what}.

Today's state-of-the-art HDD has reached a capacity of \SI{32}{TB} (over \SI{3}{TB} per disk) \cite{2024(seagatetechnology)what,2024(hernandez)high}, which is implemented by heat-assisted magnetic recording (HAMR) technology. HAMR achieves the instantaneous reduction of the coercivity of localized magnetic medium through the high temperature provided by an integrated laser \cite{2006(rottmayer)heatassisted,2008(kryder)heat,2009(challener)heatassisted,2010(stipe)magnetic}. With lower coercivity, the magnetic orientations of the medium can be reversed by available applied fields. HAMR effectively solves the writability problem of highly anisotropic materials \cite{2014(weller)hamr,2016(weller)review} thus it allows for smaller nano-grains to be employed, while maintaining the energy barrier required for long-term thermal stability \cite{1994(sharrock)time,1999(weller)thermal}.

Although HAMR has greatly increased the capacity of HDDs, the superparamagnetic limit \cite{2007(richter)transition,2012(evans)thermally,2007(piramanayagam)perpendicular} remains, making it impossible to reduce the grain size infinitely and hindering the further growth of the areal density. To break through this bottleneck, three-dimensional magnetic recording (3DMR), which utilizes more than one layer of recording medium, has been considered a very potential approach \cite{2022(roddick)new}. Based on the multi-layer recording medium, multiple levels of magnetization states can be written on one single-bit position, thereby exponentially increasing the capacity without the need for small grain size.

The implementation of 3DMR is facing a number of challenges \cite{2023(greaves)threedimensional}. The most critical of these is how to write bits sequentially and correctly to the multi-layer recording medium. A recent report \cite{2024(tozman)duallayer} substantiates the practicality of HAMR with dual-layer \ce{FePt-C} nanogranular films. In this work, we have proposed a design of a hierarchical recording architecture that features layered heat-assisted writing with a multi-head array, supporting data writing in 3DMR. Through micromagnetic modeling and simulation, we examined the architecture in a system with dual-layer \ce{FePt}-based recording medium. The field-dependent behavior of the switching possibilities for different layers of the medium was studied, and the magnetization reversal dynamics of the grains during the hierarchical recording process were elucidated. Thorough discussions on the performance and optimization of our proposed architecture were conducted, with the results demonstrating its superiority for the future realization of robust 3DMR.

\section{CONCEPT AND MODEL
\label{sec2}}
\subsection{Multi-Layer Recording Medium}
\begin{figure}[t]
    \centering
    \includegraphics[width=8.6 cm]{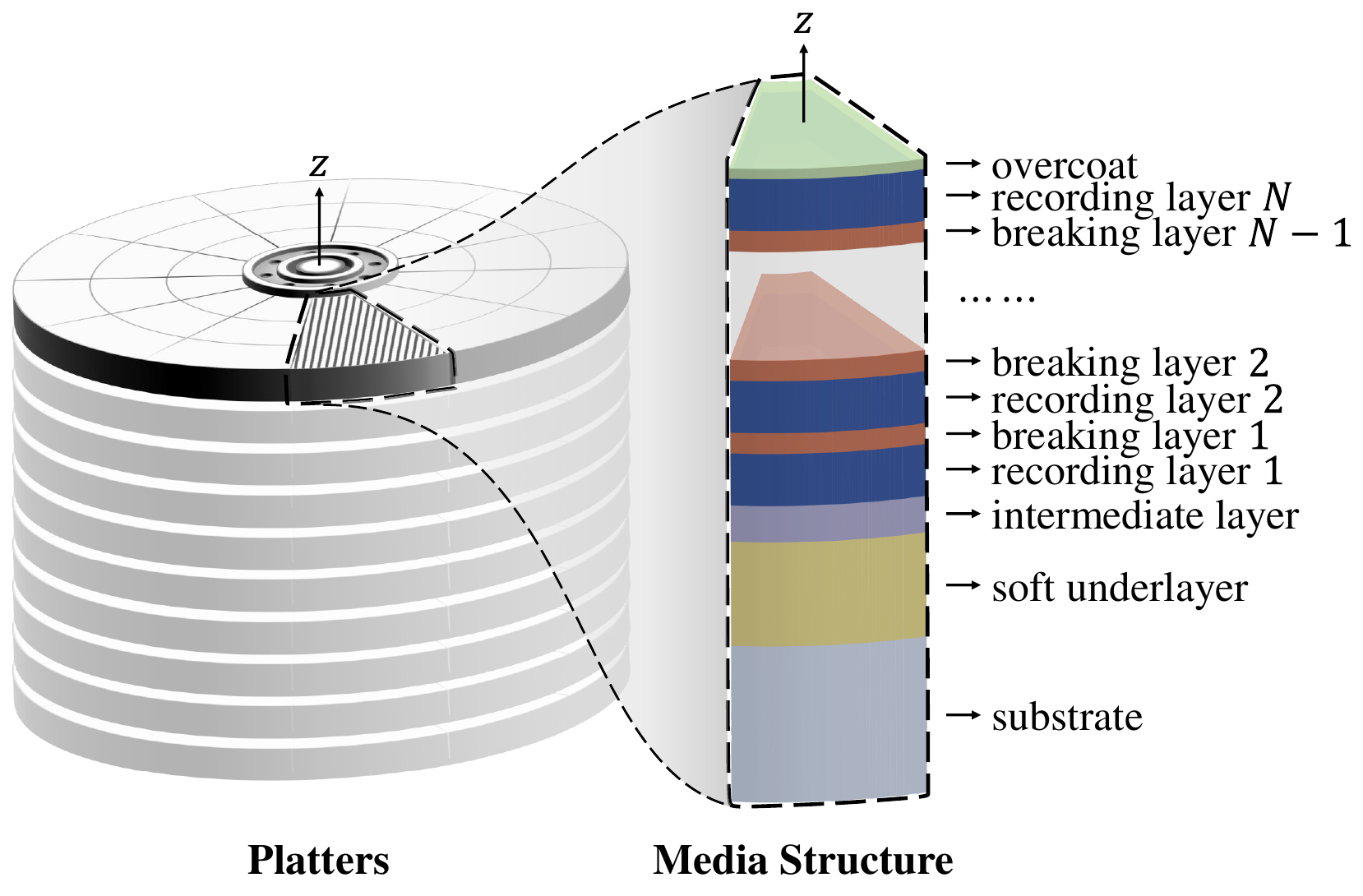}
    \caption{Structure and major components of the 3DMR media.
    \label{fig1}}
\end{figure}

An HDD usually contains several platters, each with a complex media structure. During the writing process, the alteration of magnetization occurs on a single-layer recording medium, which is made of an uniaxially anisotropic magnetic material \cite{2000(weller)high,2007(piramanayagam)perpendicular}, so existing magnetic recording is fundamentally two-dimensional. In 3DMR, there will be two or more layers used for recording. Fig.~\ref{fig1} illustrates the structure of the 3DMR media. Multiple recording layers are stacked in the perpendicular direction ($z$ axis). Two adjacent recording layers will be separated and magnetically decoupled by introducing a breaking layer between them \cite{2024(tozman)duallayer}.

\subsection{Layered Heat-Assisted Writing}
\begin{figure}[t]
    \centering
    \includegraphics[width=8.8 cm]{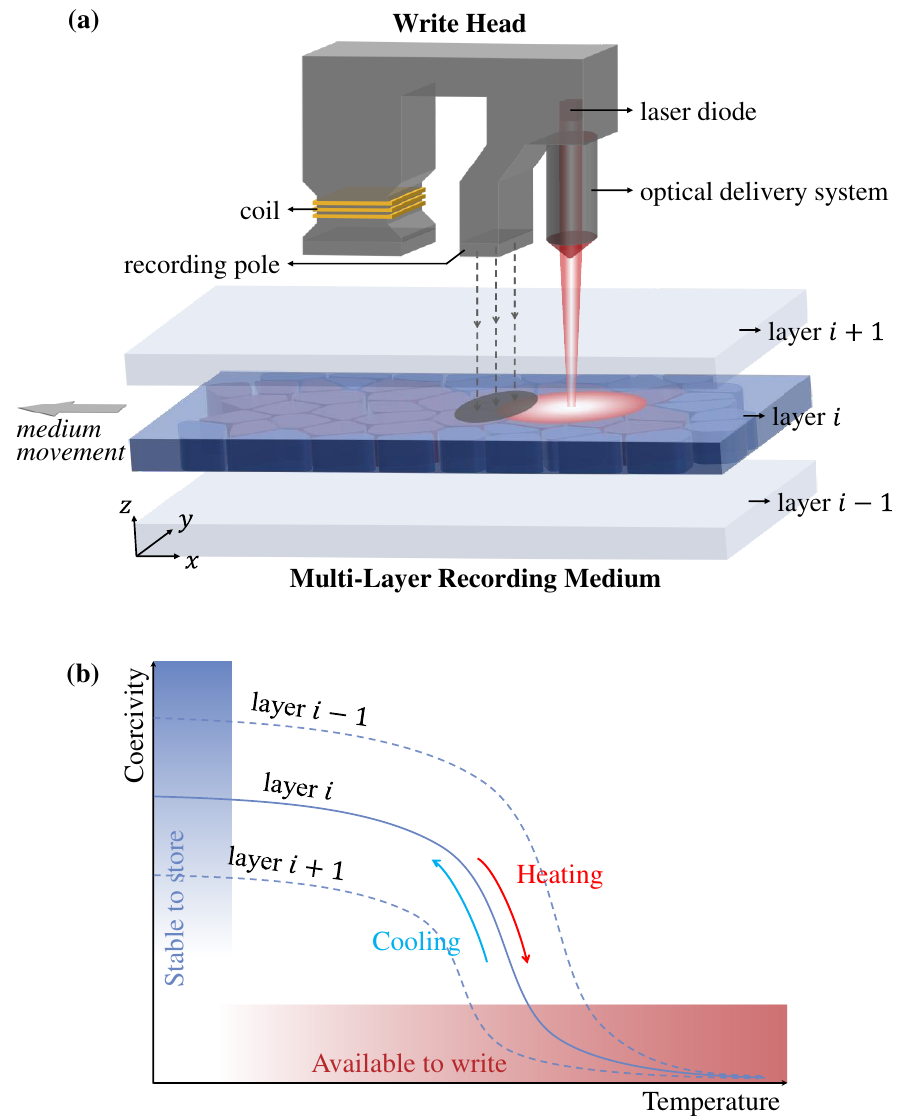}
    \caption{(a) Schematic diagram of the layered heat-assisted writing process in 3DMR. (b) Temperature dependencies of coercivity for different layers.
    \label{fig2}}
\end{figure}

To enable independent writing on either recording layer in 3DMR, it is imperative that they possess distinct anisotropy fields and can be magnetized under different physical conditions. HAMR may be a feasible way to realize 3DMR. When writing data to a specific layer, the nanoscale hot spot created by the laser and optical delivery system \cite{2009(challener)heatassisted,2010(stipe)magnetic,2015(gosciniak)novel,2016(gosciniak)novel,2020(chen)highfield,2021(chen)simple} will be perpendicularly focused on the corresponding position, locally heating the medium to near its Curie temperature and temporarily reducing its coercivity, followed by the external field from the recording pole switching the polarity of the magnetic grains during cooling, as shown in Fig.~\ref{fig2}. Adjusting the laser's focal depth and power to match the position and Curie temperature of each recording layer thus allows for separated writing in 3DMR.

\subsection{Multi-Head Array}
The distribution and gradient of the temperature are crucial factors influencing the performance of an HAMR system. The entire media must have proper heat conductivity and dissipation to ensure that data can be effectively written without being further erased \cite{2013(xu)hamr,2021(jubert)anisotropic,2021(dwivedi)graphene}. A more complex circumstance in 3DMR is that temperature variations occur not only in the $xy$ plane but also in the $z$ direction. Due to the downward conduction of the thermal field produced by the laser, the recording layer closer to the write head will inevitably have a higher temperature. As a result, when writing to the bottom layer, the data in the top layer will be erased.

\begin{figure}[t]
    \centering
    \includegraphics[width=8.8 cm]{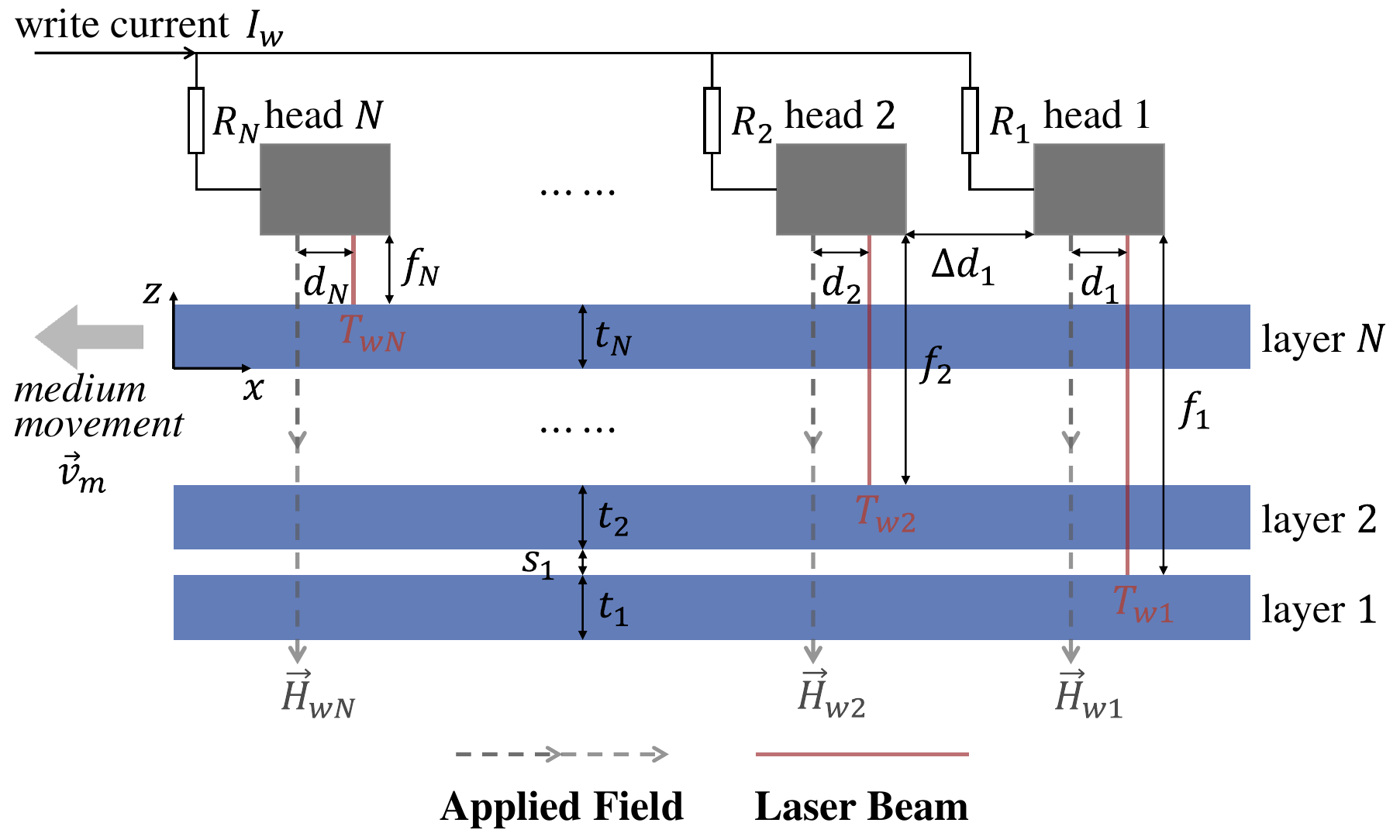}
    \caption{Schematic diagram of the hierarchical recording architecture with a multi-head array for 3DMR.
    \label{fig3}}
\end{figure}

Here we propose a hierarchical recording architecture for 3DMR, leveraging multi-pass writing with a multi-head array to sequentially write data into the multi-layer recording medium. Fig.~\ref{fig3} gives a schematic diagram of the architecture. Overall, the writing process begins from the bottom layer and progresses upward, layer by layer, as the medium moves. Each head in the array corresponds to the respective layer, i.e., head $i$ operates on layer $i$ at the $i$-th pass writing $(i = 1,2, \cdots, N)$. The laser's focal depth $f_i$ and the writing temperature $T_{wi}$ it provides, as well as the writing field $H_{wi}$, are determined by the structural position and physical properties of each recording layer, and will be either preset or controlled by the input current.

The writing fields can be independent from one another, while the writing temperatures need to satisfy the conditions:
\begin{equation}
    \left\{
\begin{aligned}
    &T_{w1} > T_{w2} > \cdots > T_{wN} \\
    &T_{wi} = T_{Ci} + \delta T_i \quad (i = 1,2, \cdots ,N)
    \label{eq1}
\end{aligned}
    \right.
\end{equation}
$T_{wi}$ and $T_{Ci}$ are the writing and Curie temperatures of head/layer $i$, respectively. $\delta T_i$ is a small positive value that represents $T_{wi}$ is slightly higher than $T_{Ci}$. It is essential to hold such a descending temperature distribution to ensure that data previously written remains unaffected in the onward process.

\subsection{Micromagnetic Model}
To simulate the writing process in our proposed hierarchical recording architecture, we have developed a micromagnetic model based on MuMax3 \cite{2014(vansteenkiste)design} and MARS \cite{2022(rannala)models}. The magnetization dynamics of grain $j$ in layer $i$ approaching the Curie temperature are investigated by solving the Landau-Lifshitz-Bloch (LLB) equation \cite{2022(rannala)models,2012(evans)stochastic}:
\begin{equation}
\begin{aligned}
    \frac{\partial \vec{m}^{i,j}}{\partial t} = &- \gamma_e \left( \vec{m}^{i,j} \times \vec{H}^{i,j}_{\rm eff} \right) \\
    &+ \frac{\gamma_e \alpha_{\parallel}}{{m^{i,j}}^2} \left( \vec{m}^{i,j} \cdot \vec{H}^{i,j}_{\rm eff} \right) \vec{m}^{i,j} \\
    &- \frac{\gamma_e \alpha_{\perp}}{{m^{i,j}}^2} \left[ \vec{m}^{i,j} \times \left( \vec{m}^{i,j} \times \left( \vec{H}^{i,j}_{\rm eff} + \zeta^{i,j}_{\perp} \right) \right) \right] \\
    &+ \zeta^{i,j}_{ad}
    \label{eq2}
\end{aligned}
\end{equation}
where $\vec{m}^{i,j} = \vec{M}^{i,j} / {M^{i,j}_s (T=0)}$ is the normalized magnetization and $m^{i,j}$ is the magnitude of $\vec{m}^{i,j}$. $\gamma_e$ is the electron gyromagnetic ratio. $\vec{H}^{i,j}_{\rm eff}$ is the effective field. $\alpha_{\parallel}$ and $\alpha_{\perp}$ are the longitudinal and transverse components of the damping of the magnetic moment, respectively, related to the Gilbert damping constant $\lambda$. $\zeta^{i,j}_{\perp}$ and $\zeta^{i,j}_{ad}$ are the diffusion coefficients that account for the thermal fluctuations, which describe the finite-temperature effects on the perpendicular and parallel components of the magnetization, respectively. In Eq.~\eqref{eq2}, the first and third terms are the precessional and damping terms for the transverse component of magnetization, similar to those in the Landau-Lifshitz-Gilbert (LLG) equation, while the second and fourth terms are introduced to account for the longitudinal relaxation of magnetization with temperature.

Our model consists of three parts: temperature, applied field, and multi-layer recording medium. Both profiles of temperature and applied field are assumed to vary continuously along the down-track direction ($x$ axis) during the whole writing process. The writing temperature of each head follows a Gaussian distribution, and the total temperature is the scalar sum of all writing temperatures:

\begin{equation}
\begin{aligned}
    T (x,t) = &\sum\limits_{i=1}^{N} \left( T_{wi} - T_{\rm env} \right) {\rm exp} \left[ \frac{- \left( x - vt - D_i \right)^2}{2 \sigma_{Ti}^2} \right] \\
    &+ T_{\rm env}
    \label{eq3}
\end{aligned}
\end{equation}
where $T_{wi}$ is the writing temperature of head $i$ and $T_{\rm env}$ is the environment temperature, taken as \SI{300}{\kelvin}. $v$ is the velocity of medium movement. $\sigma_{Ti} = {\rm FWHM}_{Ti} / \sqrt{8 \ln 2}$ is the standard deviation of the Gaussian profile of $T_{wi}$, with its full width at half maximum ${\rm FWHM}_{Ti}$. $D_i$ is used to supplement the position of $T_{wi}$ and it is calculated by:
\begin{equation}
    \left\{
\begin{aligned}
    &D_i = d_i \quad &(i = 1) \\
    &D_i = d_i + \sum\limits_{i=2}^{i} \Delta d_i \quad &(i \geq 2)
    \label{eq4}
\end{aligned}
    \right.
\end{equation}
$d_i$ is the distance between the laser and the recording pole in head $i$, indicating that the laser will be activated and interact with the medium a bit earlier. $\Delta d_i$ is the distance between the current head and the previous one, i.e., head $i$ and head $i-1$.

The writing field is considered constant within the head area, while the stray field outside the head is Gaussian distributed. Similarly, the total field is the vector sum of all writing fields:
\begin{equation}
    \vec{H} (x) = \left\{
\begin{aligned}
    &b_i \hat{u}_i H_{wi} \qquad (x \ {\rm is \ within \ the \ head \ area}) \\
    &\sum\limits_{i=1}^{N} b_i \hat{u}_i H_{wi} {\rm exp} \left[ \frac{- \left( x - X_{i} \right)^2}{2 \sigma_{Hi}^2} \right] \\
    & \qquad \qquad \quad \ (x \ {\rm is \ outside \ the \ head \ area}) 
    \label{eq5}
\end{aligned}
    \right.
\end{equation}
where $b_i = \{ 1,-1 \}$ is the current bit to be written by head $i$, $\hat{u}_i$ is a unit vector to represent the field's direction/angle of head $i$, and $H_{wi}$ is the writing field of head $i$. $\sigma_{Hi} = {\rm FWHM}_{Hi} / \sqrt{8 \ln 2}$ is analogous to $\sigma_{Ti}$ as mentioned above. $X_i$ is either the left or right edge of head $i$, depending on $x$ position. Additionally, the writing field also has a time profile $\vec{H} (t)$. The ramp time $\tau_r$ is considered when there is a change of $b_i$.

For the recording medium of each layer, \ce{FePt}-based thin film, which is a typical material of choice for the HAMR system due to its very high uniaxial anisotropy and relatively low Curie temperature \cite{2024(tozman)duallayer,2024(ogawa)exchangecoupled,2002(okamoto)chemicalorderdependent,2006(rong)sizedependent,2012(hovorka)curie,2017(liu)thermal,2017(richter)temperature}, with Voronoi grains and nonmagnetic grain boundaries is modeled. Each layer should possess distinct magnetic properties and be magnetically isolated.

\begin{table*}
    \caption{Parameters and settings of the micromagnetic model. Values such as “xxx / xxx” refer to those for the top and bottom layers, respectively. Curie temperature, uniaxial anisotropy, and grain volume of the recording medium follow a log-normal distribution. The temperature dependence of anisotropy and magnetization is described by Callen-Callen scaling \cite{1966(callen)present}: $K (T) = K (0) m (T)^{\eta}$. Here we set $\eta = 2.0$ for \ce{FePt} with 2-state anisotropy \cite{2005(mryasov)temperaturedependent}.
    \label{tbl1}}
\begin{ruledtabular}
\begin{tabular}{lc}
    Writing temperature $T_w$ (K) & 540.0 / 680.0 \\
    Distance between the laser and the recording pole in head $d$ (nm) & 1.0 \\
    Full width at half maximum of the writing temperature ${\rm FWHM}_T$ (nm) & 20.0 \\
    Head width (nm) & 20.0 \\
    Unit vector of the writing field's direction $\hat{u}_i$ & (0.0,0.0,1.0) \\
    Full width at half maximum of the stray field ${\rm FWHM}_H$ (nm) & 20.0 \\
    Ramp time of the writing field $\tau_r$ (ns) & 0.1 \\
    Saturation magnetization at \SI{0}{\kelvin} $M_s (0)$ (emc/cc) & 487.0 / 696.0 \\
    Curie temperature $T_C$ (K) & 526.0 / 620.0 \\
    Curie temperature deviation $\sigma_{TC}$ & 0.03 \\
    Uniaxial anisotropy at \SI{0}{\kelvin} $K_u (0)$ ($10^6$erg/cc) & 6.0 / 25.0 \\
    Uniaxial anisotropy deviation $\sigma_{Ku}$ & 0.15 \\
    Damping $\alpha$ & 0.1 \\
    Medium thickness $t$ (nm) & 6.0 \\
    Grain diameter $D_{G}$ (nm) & 6.0 \\
    Grain volume deviation $\sigma_{G}$ & 0.09 \\
\end{tabular}
\end{ruledtabular}
\end{table*}

Before the writing process starts, all grains are initialized to the same magnetization polarity (i.e., the recording medium is DC-erased) and the system is equilibrated at room temperature. The simulation concerns writing a given bit sequence onto a single finite-length track. During the simulation, the recording medium remains stationary, while the head array moves along the down-track direction at a velocity $v$, exerting thermal and magnetic fields to induce grain switching. Once the trailing end of the last head moves off the track, the simulation will run for a bit longer to cool the system down to room temperature and allow it to reach equilibrium again.

In our following examination, we primarily focus on the case of $N = 2$, i.e., dual-layer 3DMR with dual-head writing. Key parameters and settings are listed in Table~\ref{tbl1}.

\section{RESULTS AND DISCUSSION
\label{sec3}}
\subsection{Field-Dependent Switching Possibility}
\begin{figure}[ht]
    \centering
    \includegraphics[width=8.8 cm]{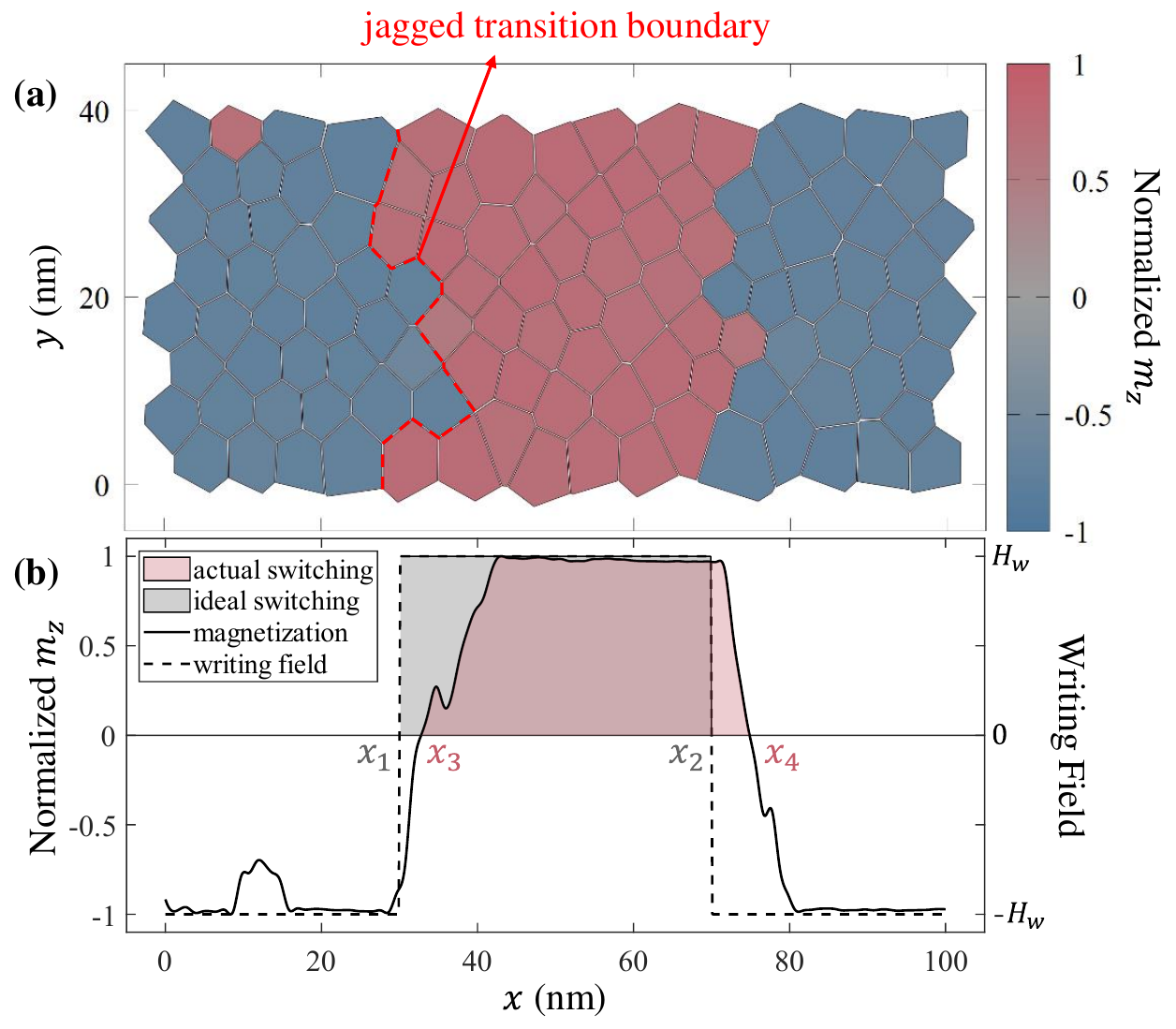}
    \caption{The results for either the top or bottom layer after individual heat-assisted writing. (a) The distribution of the $z$-axis normalized magnetization $m_z$ of grains in the $xy$ plane. (b) The $z$-axis normalized magnetization and the writing field along the down-track direction, $m_z(x)$ (solid line) and $H_w(x)$ (dashed line). Pink and gray represent the actual and ideal switching regions on the track, respectively.
    \label{fig4}}
\end{figure}

It is important for each head to be well-matched with the corresponding layer to achieve optimal writing conditions in the hierarchical recording architecture. We first performed individual simulations under the heat-assisted writing process for the top and bottom layers to determine the most proper writing field of each. The medium was initially DC-erased to all “0”s, after which a segment of “1”s of specified length was written at the center of a designated track. Fig.~\ref{fig4} shows an example of the resulting magnetization distribution in the track.

The above writing process was simulated separately for the top and bottom layers under various writing field strengths and was repeated several times to eliminate the effect of grain distribution. We calculated the effective switching possibility $SP_{\rm eff}$ through the integration of magnetization and writing field over the write window along the down-track direction (see Eq.~\eqref{eq6}), serving it as a criterion to determine whether the writing field is best-matched.
\begin{equation}
    SP_{\rm eff} = \frac{\int_{x_3}^{x_4} m_z(x) \, {\rm d}x}{\int_{x_1}^{x_2} H_w(x) \, {\rm d}x}
    \label{eq6}
\end{equation}
where $(x_2-x_1)$ and $(x_4-x_3)$ are the ideal and actual write windows, respectively. The significance of $SP_{\rm eff}$ calculation is also illustrated in Fig.~\ref{fig4}(b). The closer it is to $1$, the more accurately the data was written.

\begin{figure}[t]
    \centering
    \includegraphics[width=8.0 cm]{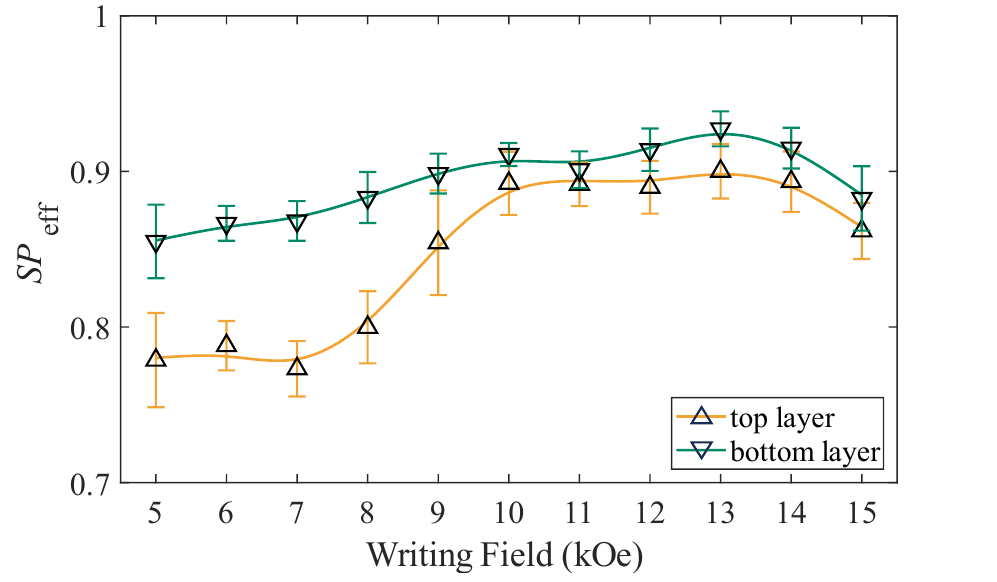}
    \caption{The dependence of effective switching possibility $SP_{\rm eff}$ on writing field for the top (yellow line) and bottom (green line) layers.
    \label{fig5}}
\end{figure}

Fig.~\ref{fig5} shows the results and indicates a strong field dependence of the calculated $SP_{\rm eff}$. This is because severe thermal fluctuations, especially near the Curie temperature, are introduced in the HAMR system \cite{2017(liu)thermal,2013(zhu)understanding,2023(everaert)impact}. Too small or too large field strength can lead to a decline in switching possibility, though for different reasons. The former arises out of incomplete writing, where some grains may fail to be magnetized to the expected orientation. The latter is caused by the erasure of neighboring bits, leading to a jagged transition edge (as shown in Fig.~\ref{fig4}(a)) and thereby broadening the transition.

Owing to the higher temperature gradient, the writing performance in the bottom layer will be better than that in the top layer. Nevertheless, the results exhibit that the optimal write fields for the top and bottom layers are \SI{13.1}{kOe} (maximum $SP_{\rm eff} = 0.900$) and \SI{13.0}{kOe} (maximum $SP_{\rm eff} = 0.927$), with very few differences. Such comparable field-dependent behavior arises because the thermal and magnetic properties of both layers in our simulation are assumed to be based on \ce{FePt}.

\subsection{Magnetization Reversal Dynamics}
To investigate the dynamical process of grain's magnetization reversal in 3DMR, the dual-head writing of consecutive transitions (square-wave binary sequence for both heads) on the dual-layer medium was simulated. We concentrated on the temporal evolution of magnetization as the two write heads pass over the grain sequentially.

Here we select two specific grains at the same position on the written track for discussion, one from the top layer and the other from the bottom layer. The simulation was performed at a write speed of \SI{5}{m/s} with a bit length of \SI{20}{nm}, and the relevant results are shown in Fig.~\ref{fig6}(a). The entire magnetization reversal during the hierarchical recording process can be divided into four stages:

\begin{figure}[t]
    \centering
    \includegraphics[width=8.8 cm]{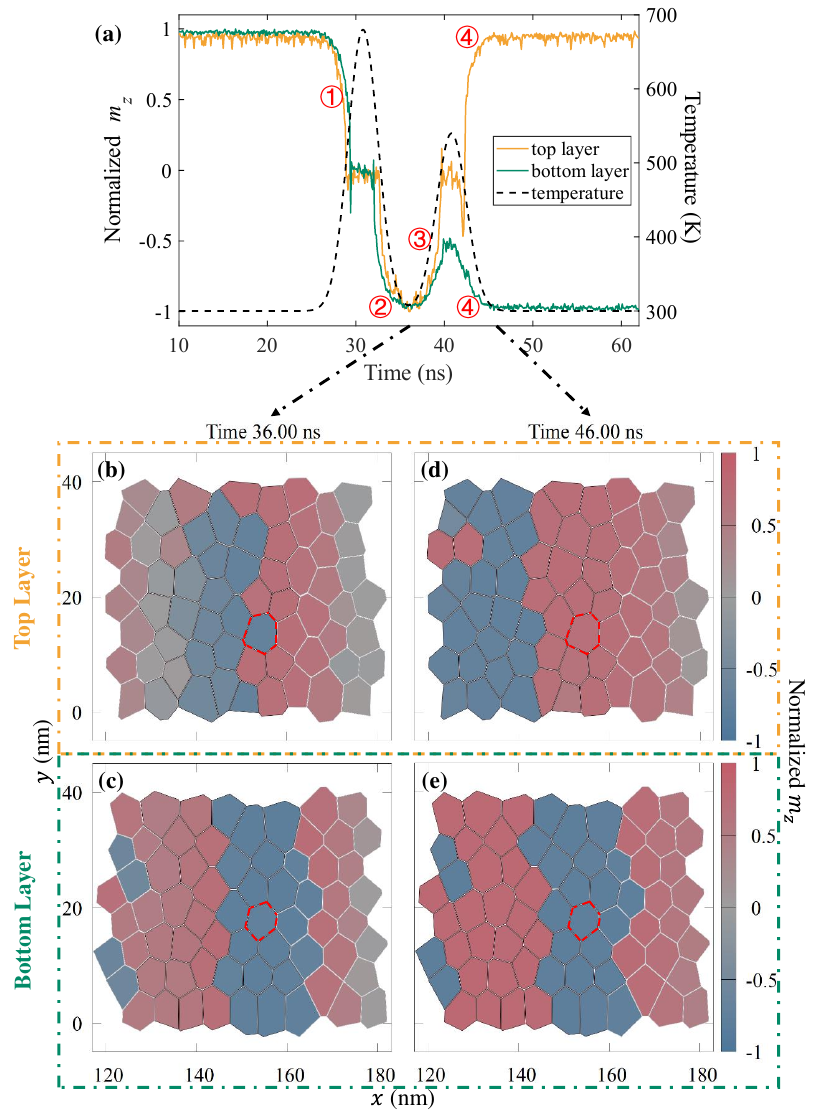}
    \caption{(a) Temporal evolution of the magnetization reversal of grains from the top (yellow line) and bottom (green line) layers during the hierarchical recording process, which is discussed in stages \textcircled{1}-\textcircled{4}. The dashed line represents the variation of the grain's temperature over time. (b-e) The distribution of the $z$-axis normalized magnetization $m_z$ of top-layer (yellow box) and bottom-layer (green box) grains in the $xy$ plane after the first-pass (b,c) and second-pass (d,e) writings. The grains shown in (a) are outlined with the red dashed lines.
    \label{fig6}}
\end{figure}

\begin{enumerate}
    \item The medium was gradually heated to the higher writing temperature, of which $T_{w1} > T_{C1} > T_{C2}$, and the coercivity of both the top and bottom layers was rapidly reduced. The grains transitioned from ferromagnetic to paramagnetic ($m_z \rightarrow 0$) as the temperature rose.
    \item As the head moved forward, the grains regained their ferromagnetic state while cooling, and the magnetization was reversed under the effect of the writing field. At this point, the bits corresponding to the top and bottom layers switched from “1” to “0”.
    \item The second-pass writing began, and the medium was heated to the lower writing temperature, of which $T_{C1} > T_{w2} > T_{C2}$. Consequently, only the magnetization of the top-layer grain would be erased. The coercivity and magnetization of the bottom-layer grain decreased as the temperature rose but did not meet the writable condition, thus the bottom-layer grain would remain in the state achieved after the first-pass writing.
    \item Finally, the magnetization of the top-layer grain was reversed by the writing field, while the bottom-layer grain remained unaffected and returned to its previous state. The medium would reach equilibrium at room temperature, with data “1” and “0” separately stored in the corresponding bits of the top and bottom layers.
\end{enumerate}

It is worth noting that during the writing process, the magnetization of the top-layer grain exhibits somewhat greater fluctuations or noise compared to the bottom-layer one. This may be attributed to the lower anisotropy and saturation magnetization of the top-layer medium, making it more sensitive to variations in magnetic and thermal fields.

Fig.~\ref{fig6}(b-e) also show the distribution of grains' magnetization after the first-pass and second-pass writings, respectively. It can be observed that the outcome for the bottom layer has been determined after the first-pass writing, while the outcome for the top layer is defined by the second-pass writing, i.e., head $1$ writes to layer $1$ and head $2$ writes to layer $2$. This effectively confirms the feasibility of our proposed hierarchical recording architecture. By controlling the current (magnitude, phase, etc.), it is possible to write any desired pattern, including “00”, “01”, “10”, and “11”, into the dual-layer medium.

\subsection{Medium Signal-to-Noise Ratio}
According to the above results, it can be found that there are unexpected switchings or abrupt transitions in magnetization for grains of both the top and bottom layers, which stem from the combined interactions of multiple factors and will introduce noise into the 3DMR system. In the hierarchical recording architecture, the distance between adjacent heads $\Delta d$ emerges as a new parameter that impacts the recording quality. Hence, we repeated the writing simulations with varying $\Delta d$, under different write speeds and bit lengths. The medium signal-to-noise ratio (SNR) was calculated under these conditions to evaluate the performance of the 3DMR system.

We define the medium SNR as dividing signal power by noise power, which are denoted by the average magnetization of transitions and its variance, respectively:
\begin{equation}
    {\rm Medium \ SNR}  = 10 \log_{10} \left[ \frac{\int_{-{\rm BL} / 2}^{{\rm BL} / 2} \overline{m_z}(x)^2 \, {\rm d}x}{\int_{-{\rm BL} / 2}^{{\rm BL} / 2} \overline{\Delta m_z}(x)^2 \, {\rm d}x} \right]
    \label{eq7}
\end{equation}
where ${\rm BL}$ is the bit length. The average and variance of the magnetization profiles were obtained by first averaging $m_z$ of all grains over a fixed width across the cross-track direction, then shifting consecutive transitions along the down-track direction to align them at transition centers.

\begin{figure*}[t]
    \includegraphics[width=17.5 cm]{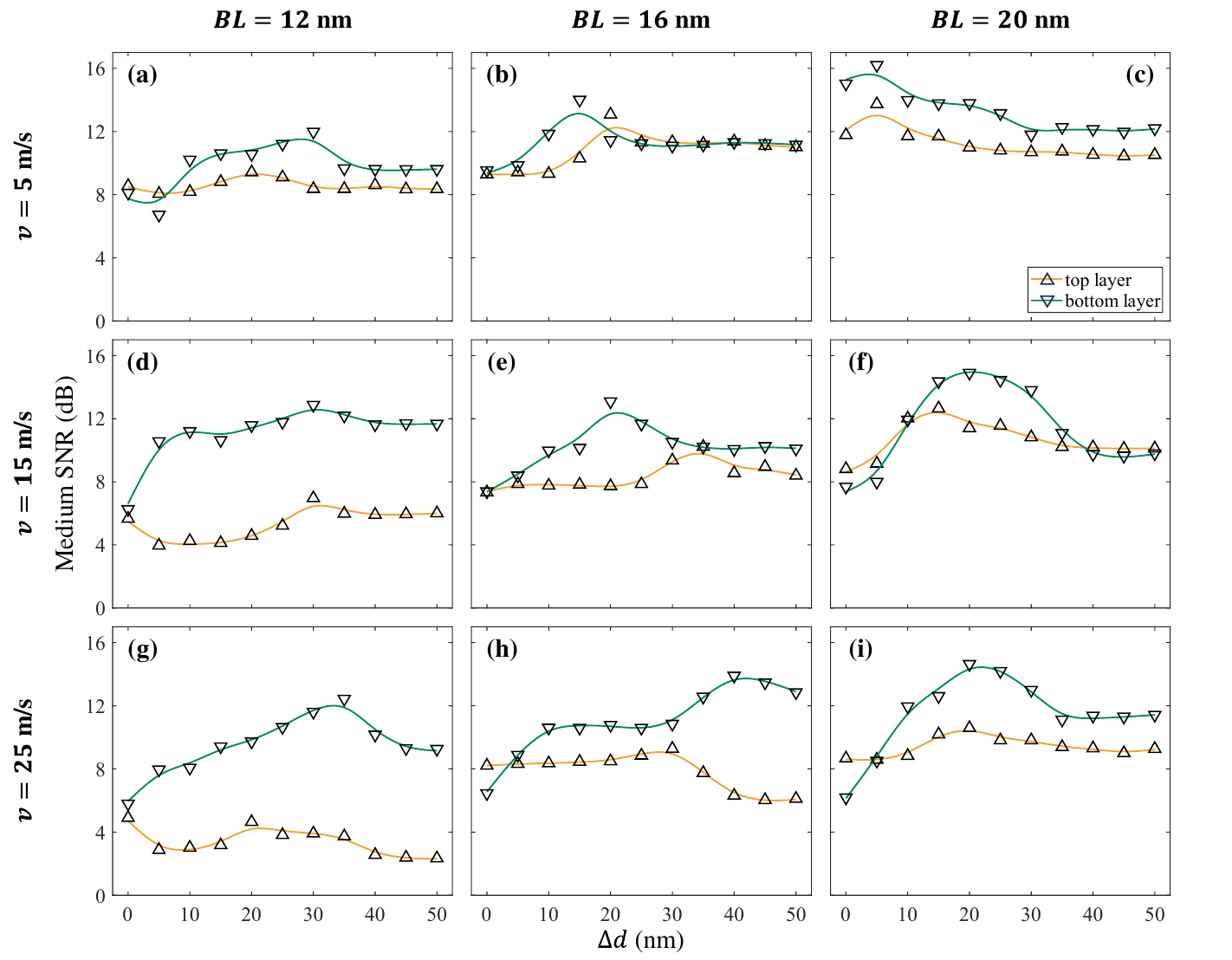}
    \caption{The medium SNR versus the head-to-head distance $\Delta d$ under different write speeds and bit lengths for the top (yellow lines) and bottom (green lines) layers.
    \label{fig7}}
\end{figure*}

The results of medium SNRs are shown in Fig.~\ref{fig7} with very intriguing trends. To begin with, as the head-to-head distance $\Delta d$ grows from zero, the medium SNR will gradually improve for both the top and bottom layers. This is due to the thermal profile generated by the dual-head structure becoming increasingly distinguishable. Therefore, the system's peak temperature decreases, while the thermal gradient at the midpoint between the two heads increases, facilitating more precise writing of their respective transition sequences. Simultaneously, the stray field at the edges of the heads diminishes, which reduces the likelihood of erroneous switching of grains in adjacent bits and thereby minimizes inter-symbol interference.

Few abnormal results are observed as well, such as instances where the SNR is unexpectedly higher when $\Delta d = 0$ (in Fig.~\ref{fig7}(a,d,g)). In this case, the two heads are perfectly aligned, and both layers can be treated as a conventional two-dimensional HAMR system performing synchronized writing. The influence from multiple heads is instead mitigated.

Interestingly, for all conditions, when $\Delta d$ continues to increase, the medium SNR experiences a slight decline before leveling off, which is more pronounced for the bottom layer. We speculate that an excessively large $\Delta d$, or the involvement of additional heads, broadens the region influenced by thermal fluctuations, introducing noise that may overshadow the gains in SNR from the enhanced thermal gradient. This is particularly true for the bottom layer since its data must endure extra perturbation during the second-pass writing. Furthermore, this suggests the existence of an optimal distance $\Delta d_{opt}$, at which the medium SNR reaches its maximum. We accordingly calculated $\Delta d_{opt}$ and the corresponding ${\rm SNR}_{max}$ for different write speeds and bit lengths. The ${\rm SNR}_{max}$ here refers to the whole system, assuming that the magnetization from the top and bottom layers contributes equally to the signal and noise powers. The results are presented in Table~\ref{tbl2}.

From Fig.~\ref{fig7} and Table~\ref{tbl2}, it can be told that similar to conventional HAMR \cite{2013(zhu)understanding,2016(vogler)basic}, a higher write speed leads to reduced thermal gradient, and a smaller bit length (i.e., a higher linear density) introduces more severe inter-symbol interference, both of which contribute to a decrease in overall SNR. In dual-layer 3DMR, the SNR of the top-layer medium is typically lower than that of the bottom-layer medium, and it is also more susceptible to noise under different conditions. The hierarchical recording architecture yields a maximum SNR of about $8$-\SI{14}{dB} for the configurations we tested. Generally, the ${\rm SNR}_{max}$ can be achieved when $\Delta d$ is averagely 2-4 times bit lengths. A higher ${\rm SNR}_{max}$ indicates stronger noise resistance in the system, allowing for a smaller $\Delta d_{opt}$, which opens up the potential for expanding to more recording layers and a larger multi-head array in future 3DMR. Trade-offs will be needed to balance the write speed, the areal density of each layer, and the total capacity. For example, while the multi-head array shares the same relative moving speed, the multi-layer medium can adopt different density combinations, which may further enhance the architecture performance.

\begin{table}[t]
    \caption{The optimal head-to-head distance $\Delta d_{opt}$ and the corresponding ${\rm SNR}_{max}$ for different write speeds and bit lengths.
    \label{tbl2}}
\begin{ruledtabular}
\begin{tabular}{cccc}
    $v$ (m/s) & ${\rm BL}$ (nm) & $\Delta d_{opt}$ (nm) & ${\rm SNR}_{max}$ (dB) \\
    \hline
    5 & 12 & 44.8 & 10.355 \\
    5 & 16 & 38.2 & 12.063 \\
    5 & 20 & 24.5 & 14.303 \\
    15 & 12 & 50.6 & 9.679 \\
    15 & 16 & 42.0 & 10.133 \\
    15 & 20 & 37.6 & 13.498 \\
    25 & 12 & 52.2 & 8.306 \\
    25 & 16 & 52.5 & 10.100 \\
    25 & 20 & 41.2 & 12.325 \\
\end{tabular}
\end{ruledtabular}
\end{table}

\section{CONCLUSION
\label{sec4}}
To address the challenge of data writing in 3DMR, in this work, we have proposed a hierarchical recording architecture that incorporates layered heat assistance and a multi-head array. Each head operates on a respective layer of the recording medium in a stepwise fashion, writing bit sequences perpendicularly from the bottom to the top, preventing data erasure caused by the downward heat conduction.

Our results from the micromagnetic simulation of a dual-layer 3DMR system with \ce{FePt}-based thin films not only successfully confirm the technical feasibility of the proposed architecture, but also reveal the magnetization reversal mechanism of the grains during the hierarchical recording process. The architecture ultimately attains impressive switching possibility and medium SNR for each layer under suitable writing conditions. In particular, it is discovered that the trend in the medium SNR is not monotonic as the head-to-head distance grows. There is an optimal distance that meets the maximum SNR. If the system is more capable of withstanding noise or interference (e.g., with a lower write speed or a larger bit length), the higher the overall SNR and the smaller this optimal distance will become. This implies that the architecture has the potential to scale 3DMR to more recording layers, offering the prospect of achieving ultra-large data storage capacity.

\begin{acknowledgments}
This work was supported by the National Natural Science Foundation of China (Grant No. 62272178) and the Fundamental Research Funds for the Central Universities (Grant No. YCJJ20242107). Y. J. acknowledges the support from the China Scholarship Council (Grant No. 202306160094).
\end{acknowledgments}

\bibliography{./bib/jian_hierarchical.v1.bib}

\begin{thebibliography}{40}%
\makeatletter
\providecommand \@ifxundefined [1]{%
 \@ifx{#1\undefined}
}%
\providecommand \@ifnum [1]{%
 \ifnum #1\expandafter \@firstoftwo
 \else \expandafter \@secondoftwo
 \fi
}%
\providecommand \@ifx [1]{%
 \ifx #1\expandafter \@firstoftwo
 \else \expandafter \@secondoftwo
 \fi
}%
\providecommand \natexlab [1]{#1}%
\providecommand \enquote  [1]{``#1''}%
\providecommand \bibnamefont  [1]{#1}%
\providecommand \bibfnamefont [1]{#1}%
\providecommand \citenamefont [1]{#1}%
\providecommand \href@noop [0]{\@secondoftwo}%
\providecommand \href [0]{\begingroup \@sanitize@url \@href}%
\providecommand \@href[1]{\@@startlink{#1}\@@href}%
\providecommand \@@href[1]{\endgroup#1\@@endlink}%
\providecommand \@sanitize@url [0]{\catcode `\\12\catcode `\$12\catcode `\&12\catcode `\#12\catcode `\^12\catcode `\_12\catcode `\%12\relax}%
\providecommand \@@startlink[1]{}%
\providecommand \@@endlink[0]{}%
\providecommand \url  [0]{\begingroup\@sanitize@url \@url }%
\providecommand \@url [1]{\endgroup\@href {#1}{\urlprefix }}%
\providecommand \urlprefix  [0]{URL }%
\providecommand \Eprint [0]{\href }%
\providecommand \doibase [0]{https://doi.org/}%
\providecommand \selectlanguage [0]{\@gobble}%
\providecommand \bibinfo  [0]{\@secondoftwo}%
\providecommand \bibfield  [0]{\@secondoftwo}%
\providecommand \translation [1]{[#1]}%
\providecommand \BibitemOpen [0]{}%
\providecommand \bibitemStop [0]{}%
\providecommand \bibitemNoStop [0]{.\EOS\space}%
\providecommand \EOS [0]{\spacefactor3000\relax}%
\providecommand \BibitemShut  [1]{\csname bibitem#1\endcsname}%
\let\auto@bib@innerbib\@empty
\bibitem [{\citenamefont {Wright}(2024)}]{2024(wright)worldwide}%
  \BibitemOpen
  \bibfield  {author} {\bibinfo {author} {\bibfnamefont {A.}~\bibnamefont {Wright}},\ }\href@noop {} {\emph {\bibinfo {title} {Worldwide {{IDC Global DataSphere Forecast}}, 2024--2028: {{AI Everywhere}}, {{But Upsurge}} in {{Data Will Take Time}}}}},\ \bibinfo {type} {Tech. Rep.}\ (\bibinfo {year} {2024})\BibitemShut {NoStop}%
\bibitem [{\citenamefont {Burns}(2024)}]{2024(burns)worldwide}%
  \BibitemOpen
  \bibfield  {author} {\bibinfo {author} {\bibfnamefont {E.}~\bibnamefont {Burns}},\ }\href@noop {} {\emph {\bibinfo {title} {Worldwide {{Hard Disk Drive Forecast}}, 2024--2028}}},\ \bibinfo {type} {Tech. Rep.}\ (\bibinfo {year} {2024})\BibitemShut {NoStop}%
\bibitem [{\citenamefont {{Seagate Technology}}(2024)}]{2024(seagatetechnology)what}%
  \BibitemOpen
  \bibfield  {author} {\bibinfo {author} {\bibnamefont {{Seagate Technology}}},\ }\href@noop {} {\emph {\bibinfo {title} {What {{Is Mozaic}} 3+, {{How Does It Work}}, and {{What Can It Do}} for {{My Data Center}}?}}},\ \bibinfo {type} {Tech. Rep.}\ (\bibinfo {year} {2024})\BibitemShut {NoStop}%
\bibitem [{\citenamefont {Hernandez}\ \emph {et~al.}(2024)\citenamefont {Hernandez}, \citenamefont {Granz},\ and\ \citenamefont {Kimble}}]{2024(hernandez)high}%
  \BibitemOpen
  \bibfield  {author} {\bibinfo {author} {\bibfnamefont {S.}~\bibnamefont {Hernandez}}, \bibinfo {author} {\bibfnamefont {S.}~\bibnamefont {Granz}},\ and\ \bibinfo {author} {\bibfnamefont {S.}~\bibnamefont {Kimble}},\ }\bibfield  {title} {\bibinfo {title} {High {{Areal Density HAMR Demonstrations}}},\ }in\ \href@noop {} {\emph {\bibinfo {booktitle} {2024 {{IEEE}} 35th Magnetic Recording Conference ({{TMRC}})}}}\ (\bibinfo {year} {2024})\ pp.\ \bibinfo {pages} {1--2}\BibitemShut {NoStop}%
\bibitem [{\citenamefont {Rottmayer}\ \emph {et~al.}(2006)\citenamefont {Rottmayer}, \citenamefont {Batra}, \citenamefont {Buechel}, \citenamefont {Challener}, \citenamefont {Hohlfeld}, \citenamefont {Kubota}, \citenamefont {Li}, \citenamefont {Lu}, \citenamefont {Mihalcea}, \citenamefont {Mountfield}, \citenamefont {Pelhos}, \citenamefont {Peng}, \citenamefont {Rausch}, \citenamefont {Seigler}, \citenamefont {Weller},\ and\ \citenamefont {Yang}}]{2006(rottmayer)heatassisted}%
  \BibitemOpen
  \bibfield  {author} {\bibinfo {author} {\bibfnamefont {R.}~\bibnamefont {Rottmayer}}, \bibinfo {author} {\bibfnamefont {S.}~\bibnamefont {Batra}}, \bibinfo {author} {\bibfnamefont {D.}~\bibnamefont {Buechel}}, \bibinfo {author} {\bibfnamefont {W.}~\bibnamefont {Challener}}, \bibinfo {author} {\bibfnamefont {J.}~\bibnamefont {Hohlfeld}}, \bibinfo {author} {\bibfnamefont {Y.}~\bibnamefont {Kubota}}, \bibinfo {author} {\bibfnamefont {L.}~\bibnamefont {Li}}, \bibinfo {author} {\bibfnamefont {B.}~\bibnamefont {Lu}}, \bibinfo {author} {\bibfnamefont {C.}~\bibnamefont {Mihalcea}}, \bibinfo {author} {\bibfnamefont {K.}~\bibnamefont {Mountfield}}, \bibinfo {author} {\bibfnamefont {K.}~\bibnamefont {Pelhos}}, \bibinfo {author} {\bibfnamefont {C.}~\bibnamefont {Peng}}, \bibinfo {author} {\bibfnamefont {T.}~\bibnamefont {Rausch}}, \bibinfo {author} {\bibfnamefont {M.}~\bibnamefont {Seigler}}, \bibinfo {author} {\bibfnamefont {D.}~\bibnamefont {Weller}},\ and\ \bibinfo {author} {\bibfnamefont {X.-M.}\ \bibnamefont {Yang}},\ }\bibfield  {title} {\bibinfo {title} {Heat-{{Assisted Magnetic Recording}}},\ }\href {https://doi.org/10.1109/TMAG.2006.879572} {\bibfield  {journal} {\bibinfo  {journal} {IEEE Trans. Magn.}\ }\textbf {\bibinfo {volume} {42}},\ \bibinfo {pages} {2417} (\bibinfo {year} {2006})}\BibitemShut {NoStop}%
\bibitem [{\citenamefont {Kryder}\ \emph {et~al.}(2008)\citenamefont {Kryder}, \citenamefont {Gage}, \citenamefont {McDaniel}, \citenamefont {Challener}, \citenamefont {Rottmayer}, \citenamefont {Ju}, \citenamefont {Hsia},\ and\ \citenamefont {Erden}}]{2008(kryder)heat}%
  \BibitemOpen
  \bibfield  {author} {\bibinfo {author} {\bibfnamefont {M.~H.}\ \bibnamefont {Kryder}}, \bibinfo {author} {\bibfnamefont {E.~C.}\ \bibnamefont {Gage}}, \bibinfo {author} {\bibfnamefont {T.~W.}\ \bibnamefont {McDaniel}}, \bibinfo {author} {\bibfnamefont {W.~A.}\ \bibnamefont {Challener}}, \bibinfo {author} {\bibfnamefont {R.~E.}\ \bibnamefont {Rottmayer}}, \bibinfo {author} {\bibfnamefont {G.}~\bibnamefont {Ju}}, \bibinfo {author} {\bibfnamefont {Y.-T.}\ \bibnamefont {Hsia}},\ and\ \bibinfo {author} {\bibfnamefont {M.~F.}\ \bibnamefont {Erden}},\ }\bibfield  {title} {\bibinfo {title} {Heat {{Assisted Magnetic Recording}}},\ }\href {https://doi.org/10.1109/JPROC.2008.2004315} {\bibfield  {journal} {\bibinfo  {journal} {Proc. IEEE}\ }\textbf {\bibinfo {volume} {96}},\ \bibinfo {pages} {1810} (\bibinfo {year} {2008})}\BibitemShut {NoStop}%
\bibitem [{\citenamefont {Challener}\ \emph {et~al.}(2009)\citenamefont {Challener}, \citenamefont {Peng}, \citenamefont {Itagi}, \citenamefont {Karns}, \citenamefont {Peng}, \citenamefont {Peng}, \citenamefont {Yang}, \citenamefont {Zhu}, \citenamefont {Gokemeijer}, \citenamefont {Hsia}, \citenamefont {Ju}, \citenamefont {Rottmayer}, \citenamefont {Seigler},\ and\ \citenamefont {Gage}}]{2009(challener)heatassisted}%
  \BibitemOpen
  \bibfield  {author} {\bibinfo {author} {\bibfnamefont {W.~A.}\ \bibnamefont {Challener}}, \bibinfo {author} {\bibfnamefont {C.}~\bibnamefont {Peng}}, \bibinfo {author} {\bibfnamefont {A.~V.}\ \bibnamefont {Itagi}}, \bibinfo {author} {\bibfnamefont {D.}~\bibnamefont {Karns}}, \bibinfo {author} {\bibfnamefont {W.}~\bibnamefont {Peng}}, \bibinfo {author} {\bibfnamefont {Y.}~\bibnamefont {Peng}}, \bibinfo {author} {\bibfnamefont {X.}~\bibnamefont {Yang}}, \bibinfo {author} {\bibfnamefont {X.}~\bibnamefont {Zhu}}, \bibinfo {author} {\bibfnamefont {N.~J.}\ \bibnamefont {Gokemeijer}}, \bibinfo {author} {\bibfnamefont {Y.-T.}\ \bibnamefont {Hsia}}, \bibinfo {author} {\bibfnamefont {G.}~\bibnamefont {Ju}}, \bibinfo {author} {\bibfnamefont {R.~E.}\ \bibnamefont {Rottmayer}}, \bibinfo {author} {\bibfnamefont {M.~A.}\ \bibnamefont {Seigler}},\ and\ \bibinfo {author} {\bibfnamefont {E.~C.}\ \bibnamefont {Gage}},\ }\bibfield  {title} {\bibinfo {title} {Heat-assisted magnetic recording by a near-field transducer with efficient optical energy transfer},\ }\href {https://doi.org/10.1038/nphoton.2009.26} {\bibfield  {journal} {\bibinfo  {journal} {Nat. Photonics}\ }\textbf {\bibinfo {volume} {3}},\ \bibinfo {pages} {220} (\bibinfo {year} {2009})}\BibitemShut {NoStop}%
\bibitem [{\citenamefont {Stipe}\ \emph {et~al.}(2010)\citenamefont {Stipe}, \citenamefont {Strand}, \citenamefont {Poon}, \citenamefont {Balamane}, \citenamefont {Boone}, \citenamefont {Katine}, \citenamefont {Li}, \citenamefont {Rawat}, \citenamefont {Nemoto}, \citenamefont {Hirotsune}, \citenamefont {Hellwig}, \citenamefont {Ruiz}, \citenamefont {Dobisz}, \citenamefont {Kercher}, \citenamefont {Robertson}, \citenamefont {Albrecht},\ and\ \citenamefont {Terris}}]{2010(stipe)magnetic}%
  \BibitemOpen
  \bibfield  {author} {\bibinfo {author} {\bibfnamefont {B.~C.}\ \bibnamefont {Stipe}}, \bibinfo {author} {\bibfnamefont {T.~C.}\ \bibnamefont {Strand}}, \bibinfo {author} {\bibfnamefont {C.~C.}\ \bibnamefont {Poon}}, \bibinfo {author} {\bibfnamefont {H.}~\bibnamefont {Balamane}}, \bibinfo {author} {\bibfnamefont {T.~D.}\ \bibnamefont {Boone}}, \bibinfo {author} {\bibfnamefont {J.~A.}\ \bibnamefont {Katine}}, \bibinfo {author} {\bibfnamefont {J.-L.}\ \bibnamefont {Li}}, \bibinfo {author} {\bibfnamefont {V.}~\bibnamefont {Rawat}}, \bibinfo {author} {\bibfnamefont {H.}~\bibnamefont {Nemoto}}, \bibinfo {author} {\bibfnamefont {A.}~\bibnamefont {Hirotsune}}, \bibinfo {author} {\bibfnamefont {O.}~\bibnamefont {Hellwig}}, \bibinfo {author} {\bibfnamefont {R.}~\bibnamefont {Ruiz}}, \bibinfo {author} {\bibfnamefont {E.}~\bibnamefont {Dobisz}}, \bibinfo {author} {\bibfnamefont {D.~S.}\ \bibnamefont {Kercher}}, \bibinfo {author} {\bibfnamefont {N.}~\bibnamefont {Robertson}}, \bibinfo {author} {\bibfnamefont {T.~R.}\ \bibnamefont {Albrecht}},\ and\ \bibinfo {author} {\bibfnamefont {B.~D.}\ \bibnamefont {Terris}},\ }\bibfield  {title} {\bibinfo {title} {Magnetic recording at 1.5 {{Pb}} m\textsuperscript{-2} using an integrated plasmonic antenna},\ }\href {https://doi.org/10.1038/nphoton.2010.90} {\bibfield  {journal} {\bibinfo  {journal} {Nat. Photonics}\ }\textbf {\bibinfo {volume} {4}},\ \bibinfo {pages} {484} (\bibinfo {year} {2010})}\BibitemShut {NoStop}%
\bibitem [{\citenamefont {Weller}\ \emph {et~al.}(2014)\citenamefont {Weller}, \citenamefont {Parker}, \citenamefont {Mosendz}, \citenamefont {Champion}, \citenamefont {Stipe}, \citenamefont {Wang}, \citenamefont {Klemmer}, \citenamefont {Ju},\ and\ \citenamefont {Ajan}}]{2014(weller)hamr}%
  \BibitemOpen
  \bibfield  {author} {\bibinfo {author} {\bibfnamefont {D.}~\bibnamefont {Weller}}, \bibinfo {author} {\bibfnamefont {G.}~\bibnamefont {Parker}}, \bibinfo {author} {\bibfnamefont {O.}~\bibnamefont {Mosendz}}, \bibinfo {author} {\bibfnamefont {E.}~\bibnamefont {Champion}}, \bibinfo {author} {\bibfnamefont {B.}~\bibnamefont {Stipe}}, \bibinfo {author} {\bibfnamefont {X.}~\bibnamefont {Wang}}, \bibinfo {author} {\bibfnamefont {T.}~\bibnamefont {Klemmer}}, \bibinfo {author} {\bibfnamefont {G.}~\bibnamefont {Ju}},\ and\ \bibinfo {author} {\bibfnamefont {A.}~\bibnamefont {Ajan}},\ }\bibfield  {title} {\bibinfo {title} {A {{HAMR Media Technology Roadmap}} to an {{Areal Density}} of 4 {{Tb}}/in\textsuperscript{2}},\ }\href {https://doi.org/10.1109/TMAG.2013.2281027} {\bibfield  {journal} {\bibinfo  {journal} {IEEE Trans. Magn.}\ }\textbf {\bibinfo {volume} {50}},\ \bibinfo {pages} {1} (\bibinfo {year} {2014})}\BibitemShut {NoStop}%
\bibitem [{\citenamefont {Weller}\ \emph {et~al.}(2016)\citenamefont {Weller}, \citenamefont {Parker}, \citenamefont {Mosendz}, \citenamefont {Lyberatos}, \citenamefont {Mitin}, \citenamefont {Safonova},\ and\ \citenamefont {Albrecht}}]{2016(weller)review}%
  \BibitemOpen
  \bibfield  {author} {\bibinfo {author} {\bibfnamefont {D.}~\bibnamefont {Weller}}, \bibinfo {author} {\bibfnamefont {G.}~\bibnamefont {Parker}}, \bibinfo {author} {\bibfnamefont {O.}~\bibnamefont {Mosendz}}, \bibinfo {author} {\bibfnamefont {A.}~\bibnamefont {Lyberatos}}, \bibinfo {author} {\bibfnamefont {D.}~\bibnamefont {Mitin}}, \bibinfo {author} {\bibfnamefont {N.~Y.}\ \bibnamefont {Safonova}},\ and\ \bibinfo {author} {\bibfnamefont {M.}~\bibnamefont {Albrecht}},\ }\bibfield  {title} {\bibinfo {title} {Review {{Article}}: {{FePt}} heat assisted magnetic recording media},\ }\href {https://doi.org/10.1116/1.4965980} {\bibfield  {journal} {\bibinfo  {journal} {J. Vac. Sci. Technol. B}\ }\textbf {\bibinfo {volume} {34}},\ \bibinfo {pages} {060801} (\bibinfo {year} {2016})}\BibitemShut {NoStop}%
\bibitem [{\citenamefont {Sharrock}(1994)}]{1994(sharrock)time}%
  \BibitemOpen
  \bibfield  {author} {\bibinfo {author} {\bibfnamefont {M.~P.}\ \bibnamefont {Sharrock}},\ }\bibfield  {title} {\bibinfo {title} {Time dependence of switching fields in magnetic recording media},\ }\href {https://doi.org/10.1063/1.358282} {\bibfield  {journal} {\bibinfo  {journal} {J. Appl. Phys.}\ }\textbf {\bibinfo {volume} {76}},\ \bibinfo {pages} {6413} (\bibinfo {year} {1994})}\BibitemShut {NoStop}%
\bibitem [{\citenamefont {Weller}\ and\ \citenamefont {Moser}(1999)}]{1999(weller)thermal}%
  \BibitemOpen
  \bibfield  {author} {\bibinfo {author} {\bibfnamefont {D.}~\bibnamefont {Weller}}\ and\ \bibinfo {author} {\bibfnamefont {A.}~\bibnamefont {Moser}},\ }\bibfield  {title} {\bibinfo {title} {Thermal effect limits in ultrahigh-density magnetic recording},\ }\href {https://doi.org/10.1109/20.809134} {\bibfield  {journal} {\bibinfo  {journal} {IEEE Trans. Magn.}\ }\textbf {\bibinfo {volume} {35}},\ \bibinfo {pages} {4423} (\bibinfo {year} {1999})}\BibitemShut {NoStop}%
\bibitem [{\citenamefont {Richter}(2007)}]{2007(richter)transition}%
  \BibitemOpen
  \bibfield  {author} {\bibinfo {author} {\bibfnamefont {H.~J.}\ \bibnamefont {Richter}},\ }\bibfield  {title} {\bibinfo {title} {The transition from longitudinal to perpendicular recording},\ }\href {https://doi.org/10.1088/0022-3727/40/9/R01} {\bibfield  {journal} {\bibinfo  {journal} {J. Phys. D: Appl. Phys.}\ }\textbf {\bibinfo {volume} {40}},\ \bibinfo {pages} {R149} (\bibinfo {year} {2007})}\BibitemShut {NoStop}%
\bibitem [{\citenamefont {Evans}\ \emph {et~al.}(2012{\natexlab{a}})\citenamefont {Evans}, \citenamefont {Chantrell}, \citenamefont {Nowak}, \citenamefont {Lyberatos},\ and\ \citenamefont {Richter}}]{2012(evans)thermally}%
  \BibitemOpen
  \bibfield  {author} {\bibinfo {author} {\bibfnamefont {R.~F.~L.}\ \bibnamefont {Evans}}, \bibinfo {author} {\bibfnamefont {R.~W.}\ \bibnamefont {Chantrell}}, \bibinfo {author} {\bibfnamefont {U.}~\bibnamefont {Nowak}}, \bibinfo {author} {\bibfnamefont {A.}~\bibnamefont {Lyberatos}},\ and\ \bibinfo {author} {\bibfnamefont {H.-J.}\ \bibnamefont {Richter}},\ }\bibfield  {title} {\bibinfo {title} {Thermally induced error: {{Density}} limit for magnetic data storage},\ }\href {https://doi.org/10.1063/1.3691196} {\bibfield  {journal} {\bibinfo  {journal} {Appl. Phys. Lett.}\ }\textbf {\bibinfo {volume} {100}},\ \bibinfo {pages} {102402} (\bibinfo {year} {2012}{\natexlab{a}})}\BibitemShut {NoStop}%
\bibitem [{\citenamefont {Piramanayagam}(2007)}]{2007(piramanayagam)perpendicular}%
  \BibitemOpen
  \bibfield  {author} {\bibinfo {author} {\bibfnamefont {S.~N.}\ \bibnamefont {Piramanayagam}},\ }\bibfield  {title} {\bibinfo {title} {Perpendicular recording media for hard disk drives},\ }\href {https://doi.org/10.1063/1.2750414} {\bibfield  {journal} {\bibinfo  {journal} {J. Appl. Phys.}\ }\textbf {\bibinfo {volume} {102}},\ \bibinfo {pages} {011301} (\bibinfo {year} {2007})},\ \Eprint {https://arxiv.org/abs/https://doi.org/10.1063/1.2750414} {https://doi.org/10.1063/1.2750414} \BibitemShut {NoStop}%
\bibitem [{\citenamefont {Roddick}\ \emph {et~al.}(2022)\citenamefont {Roddick}, \citenamefont {Kief},\ and\ \citenamefont {Takeo}}]{2022(roddick)new}%
  \BibitemOpen
  \bibfield  {author} {\bibinfo {author} {\bibfnamefont {E.}~\bibnamefont {Roddick}}, \bibinfo {author} {\bibfnamefont {M.}~\bibnamefont {Kief}},\ and\ \bibinfo {author} {\bibfnamefont {A.}~\bibnamefont {Takeo}},\ }\bibfield  {title} {\bibinfo {title} {A new {{Advanced Storage Research Consortium HDD Technology Roadmap}}},\ }in\ \href {https://doi.org/10.1109/TMRC56419.2022.9918580} {\emph {\bibinfo {booktitle} {2022 {{IEEE}} 33rd {{Magnetic Recording Conference}} ({{TMRC}})}}}\ (\bibinfo {year} {2022})\ pp.\ \bibinfo {pages} {1--2}\BibitemShut {NoStop}%
\bibitem [{\citenamefont {Greaves}(2023)}]{2023(greaves)threedimensional}%
  \BibitemOpen
  \bibfield  {author} {\bibinfo {author} {\bibfnamefont {S.}~\bibnamefont {Greaves}},\ }\bibfield  {title} {\bibinfo {title} {Three-dimensional magnetic recording},\ }\href {https://doi.org/10.1016/j.jmmm.2023.171343} {\bibfield  {journal} {\bibinfo  {journal} {Journal of Magnetism and Magnetic Materials}\ }\textbf {\bibinfo {volume} {588}},\ \bibinfo {pages} {171343} (\bibinfo {year} {2023})}\BibitemShut {NoStop}%
\bibitem [{\citenamefont {Tozman}\ \emph {et~al.}(2024)\citenamefont {Tozman}, \citenamefont {Isogami}, \citenamefont {Suzuki}, \citenamefont {Bolyachkin}, \citenamefont {{Sepehri-Amin}}, \citenamefont {Greaves}, \citenamefont {Suto}, \citenamefont {Sasaki}, \citenamefont {Chang}, \citenamefont {Kubota}, \citenamefont {Steiner}, \citenamefont {Huang}, \citenamefont {Hono},\ and\ \citenamefont {Takahashi}}]{2024(tozman)duallayer}%
  \BibitemOpen
  \bibfield  {author} {\bibinfo {author} {\bibfnamefont {P.}~\bibnamefont {Tozman}}, \bibinfo {author} {\bibfnamefont {S.}~\bibnamefont {Isogami}}, \bibinfo {author} {\bibfnamefont {I.}~\bibnamefont {Suzuki}}, \bibinfo {author} {\bibfnamefont {A.}~\bibnamefont {Bolyachkin}}, \bibinfo {author} {\bibfnamefont {H.}~\bibnamefont {{Sepehri-Amin}}}, \bibinfo {author} {\bibfnamefont {S.}~\bibnamefont {Greaves}}, \bibinfo {author} {\bibfnamefont {H.}~\bibnamefont {Suto}}, \bibinfo {author} {\bibfnamefont {Y.}~\bibnamefont {Sasaki}}, \bibinfo {author} {\bibfnamefont {T.}~\bibnamefont {Chang}}, \bibinfo {author} {\bibfnamefont {Y.}~\bibnamefont {Kubota}}, \bibinfo {author} {\bibfnamefont {P.}~\bibnamefont {Steiner}}, \bibinfo {author} {\bibfnamefont {P.}~\bibnamefont {Huang}}, \bibinfo {author} {\bibfnamefont {K.}~\bibnamefont {Hono}},\ and\ \bibinfo {author} {\bibfnamefont {Y.}~\bibnamefont {Takahashi}},\ }\bibfield  {title} {\bibinfo {title} {Dual-layer {{FePt-C}} granular media for multi-level heat-assisted magnetic recording},\ }\href {https://doi.org/10.1016/j.actamat.2024.119869} {\bibfield  {journal} {\bibinfo  {journal} {Acta Mater.}\ }\textbf {\bibinfo {volume} {271}},\ \bibinfo {pages} {119869} (\bibinfo {year} {2024})}\BibitemShut {NoStop}%
\bibitem [{\citenamefont {Weller}\ \emph {et~al.}(2000)\citenamefont {Weller}, \citenamefont {Moser}, \citenamefont {Folks}, \citenamefont {Best}, \citenamefont {Lee}, \citenamefont {Toney}, \citenamefont {Schwickert}, \citenamefont {Thiele},\ and\ \citenamefont {Doerner}}]{2000(weller)high}%
  \BibitemOpen
  \bibfield  {author} {\bibinfo {author} {\bibfnamefont {D.}~\bibnamefont {Weller}}, \bibinfo {author} {\bibfnamefont {A.}~\bibnamefont {Moser}}, \bibinfo {author} {\bibfnamefont {L.}~\bibnamefont {Folks}}, \bibinfo {author} {\bibfnamefont {M.~E.}\ \bibnamefont {Best}}, \bibinfo {author} {\bibfnamefont {W.}~\bibnamefont {Lee}}, \bibinfo {author} {\bibfnamefont {M.~F.}\ \bibnamefont {Toney}}, \bibinfo {author} {\bibfnamefont {M.}~\bibnamefont {Schwickert}}, \bibinfo {author} {\bibfnamefont {J.-U.}\ \bibnamefont {Thiele}},\ and\ \bibinfo {author} {\bibfnamefont {M.~F.}\ \bibnamefont {Doerner}},\ }\bibfield  {title} {\bibinfo {title} {High {{K\textsubscript{u} Materials Approach}} to 100 {{Gbits}}/in\textsuperscript{2}},\ }\href {https://doi.org/10.1109/20.824418} {\bibfield  {journal} {\bibinfo  {journal} {IEEE Trans. Magn.}\ }\textbf {\bibinfo {volume} {36}},\ \bibinfo {pages} {10} (\bibinfo {year} {2000})}\BibitemShut {NoStop}%
\bibitem [{\citenamefont {Gosciniak}\ \emph {et~al.}(2015)\citenamefont {Gosciniak}, \citenamefont {Mooney}, \citenamefont {Gubbins},\ and\ \citenamefont {Corbett}}]{2015(gosciniak)novel}%
  \BibitemOpen
  \bibfield  {author} {\bibinfo {author} {\bibfnamefont {J.}~\bibnamefont {Gosciniak}}, \bibinfo {author} {\bibfnamefont {M.}~\bibnamefont {Mooney}}, \bibinfo {author} {\bibfnamefont {M.}~\bibnamefont {Gubbins}},\ and\ \bibinfo {author} {\bibfnamefont {B.}~\bibnamefont {Corbett}},\ }\bibfield  {title} {\bibinfo {title} {Novel droplet near-field transducer for heat-assisted magnetic recording},\ }\href {https://doi.org/10.1515/nanoph-2015-0031} {\bibfield  {journal} {\bibinfo  {journal} {Nat. Photonics}\ }\textbf {\bibinfo {volume} {4}},\ \bibinfo {pages} {503} (\bibinfo {year} {2015})}\BibitemShut {NoStop}%
\bibitem [{\citenamefont {Gosciniak}\ \emph {et~al.}(2016)\citenamefont {Gosciniak}, \citenamefont {Mooney}, \citenamefont {Gubbinsi},\ and\ \citenamefont {Corbett}}]{2016(gosciniak)novel}%
  \BibitemOpen
  \bibfield  {author} {\bibinfo {author} {\bibfnamefont {J.}~\bibnamefont {Gosciniak}}, \bibinfo {author} {\bibfnamefont {M.}~\bibnamefont {Mooney}}, \bibinfo {author} {\bibfnamefont {M.}~\bibnamefont {Gubbinsi}},\ and\ \bibinfo {author} {\bibfnamefont {B.}~\bibnamefont {Corbett}},\ }\bibfield  {title} {\bibinfo {title} {Novel {{Mach}}--{{Zehnder Interferometer Waveguide Design}} as a {{Light Delivery System}} for {{Heat-Assisted Magnetic Recording}}},\ }\href {https://doi.org/10.1109/TMAG.2015.2477434} {\bibfield  {journal} {\bibinfo  {journal} {IEEE Trans. Magn.}\ }\textbf {\bibinfo {volume} {52}},\ \bibinfo {pages} {1} (\bibinfo {year} {2016})}\BibitemShut {NoStop}%
\bibitem [{\citenamefont {Chen}\ \emph {et~al.}(2020)\citenamefont {Chen}, \citenamefont {Chen}, \citenamefont {Gan}, \citenamefont {Luo}, \citenamefont {Huang},\ and\ \citenamefont {Lu}}]{2020(chen)highfield}%
  \BibitemOpen
  \bibfield  {author} {\bibinfo {author} {\bibfnamefont {W.}~\bibnamefont {Chen}}, \bibinfo {author} {\bibfnamefont {J.}~\bibnamefont {Chen}}, \bibinfo {author} {\bibfnamefont {Z.}~\bibnamefont {Gan}}, \bibinfo {author} {\bibfnamefont {K.}~\bibnamefont {Luo}}, \bibinfo {author} {\bibfnamefont {Z.}~\bibnamefont {Huang}},\ and\ \bibinfo {author} {\bibfnamefont {P.}~\bibnamefont {Lu}},\ }\bibfield  {title} {\bibinfo {title} {High-{{Field Enhancement}} of {{Plasmonics Antenna Using Ring Resonator}} for {{HAMR}}},\ }\href {https://doi.org/10.1109/TMAG.2020.2990525} {\bibfield  {journal} {\bibinfo  {journal} {IEEE Trans. Magn.}\ }\textbf {\bibinfo {volume} {56}},\ \bibinfo {pages} {1} (\bibinfo {year} {2020})}\BibitemShut {NoStop}%
\bibitem [{\citenamefont {Chen}\ \emph {et~al.}(2021)\citenamefont {Chen}, \citenamefont {Chen}, \citenamefont {Gan}, \citenamefont {Ma}, \citenamefont {Luo}, \citenamefont {Huang}, \citenamefont {He},\ and\ \citenamefont {Lu}}]{2021(chen)simple}%
  \BibitemOpen
  \bibfield  {author} {\bibinfo {author} {\bibfnamefont {W.}~\bibnamefont {Chen}}, \bibinfo {author} {\bibfnamefont {J.}~\bibnamefont {Chen}}, \bibinfo {author} {\bibfnamefont {Z.}~\bibnamefont {Gan}}, \bibinfo {author} {\bibfnamefont {Y.}~\bibnamefont {Ma}}, \bibinfo {author} {\bibfnamefont {K.}~\bibnamefont {Luo}}, \bibinfo {author} {\bibfnamefont {Z.}~\bibnamefont {Huang}}, \bibinfo {author} {\bibfnamefont {Y.}~\bibnamefont {He}},\ and\ \bibinfo {author} {\bibfnamefont {P.}~\bibnamefont {Lu}},\ }\bibfield  {title} {\bibinfo {title} {A simple and effective semi-circle resonator system for bit-patterned {{HAMR}}},\ }\href {https://doi.org/10.1016/j.physleta.2020.127129} {\bibfield  {journal} {\bibinfo  {journal} {Phys. Lett. A}\ }\textbf {\bibinfo {volume} {391}},\ \bibinfo {pages} {127129} (\bibinfo {year} {2021})}\BibitemShut {NoStop}%
\bibitem [{\citenamefont {Xu}\ \emph {et~al.}(2013)\citenamefont {Xu}, \citenamefont {Cen}, \citenamefont {Goh}, \citenamefont {Li}, \citenamefont {Ye}, \citenamefont {Zhang}, \citenamefont {Yang}, \citenamefont {Toh},\ and\ \citenamefont {Quan}}]{2013(xu)hamr}%
  \BibitemOpen
  \bibfield  {author} {\bibinfo {author} {\bibfnamefont {B.}~\bibnamefont {Xu}}, \bibinfo {author} {\bibfnamefont {Z.}~\bibnamefont {Cen}}, \bibinfo {author} {\bibfnamefont {J.~H.}\ \bibnamefont {Goh}}, \bibinfo {author} {\bibfnamefont {J.}~\bibnamefont {Li}}, \bibinfo {author} {\bibfnamefont {K.}~\bibnamefont {Ye}}, \bibinfo {author} {\bibfnamefont {J.}~\bibnamefont {Zhang}}, \bibinfo {author} {\bibfnamefont {H.}~\bibnamefont {Yang}}, \bibinfo {author} {\bibfnamefont {Y.~T.}\ \bibnamefont {Toh}},\ and\ \bibinfo {author} {\bibfnamefont {C.}~\bibnamefont {Quan}},\ }\bibfield  {title} {\bibinfo {title} {{{HAMR Media Design}} in {{Optical}} and {{Thermal Aspects}}},\ }\href {https://doi.org/10.1109/TMAG.2013.2257703} {\bibfield  {journal} {\bibinfo  {journal} {IEEE Trans. Magn.}\ }\textbf {\bibinfo {volume} {49}},\ \bibinfo {pages} {2559} (\bibinfo {year} {2013})}\BibitemShut {NoStop}%
\bibitem [{\citenamefont {Jubert}\ \emph {et~al.}(2021)\citenamefont {Jubert}, \citenamefont {Santos}, \citenamefont {Le}, \citenamefont {Ozdol},\ and\ \citenamefont {Papusoi}}]{2021(jubert)anisotropic}%
  \BibitemOpen
  \bibfield  {author} {\bibinfo {author} {\bibfnamefont {P.-O.}\ \bibnamefont {Jubert}}, \bibinfo {author} {\bibfnamefont {T.}~\bibnamefont {Santos}}, \bibinfo {author} {\bibfnamefont {T.}~\bibnamefont {Le}}, \bibinfo {author} {\bibfnamefont {B.}~\bibnamefont {Ozdol}},\ and\ \bibinfo {author} {\bibfnamefont {C.}~\bibnamefont {Papusoi}},\ }\bibfield  {title} {\bibinfo {title} {Anisotropic {{Heatsinks}} for {{Heat-Assisted Magnetic Recording}}},\ }\href {https://doi.org/10.1109/TMAG.2020.3019802} {\bibfield  {journal} {\bibinfo  {journal} {IEEE Trans. Magn.}\ }\textbf {\bibinfo {volume} {57}},\ \bibinfo {pages} {1} (\bibinfo {year} {2021})}\BibitemShut {NoStop}%
\bibitem [{\citenamefont {Dwivedi}\ \emph {et~al.}(2021)\citenamefont {Dwivedi}, \citenamefont {Ott}, \citenamefont {Sasikumar}, \citenamefont {Dou}, \citenamefont {Yeo}, \citenamefont {Narayanan}, \citenamefont {Sassi}, \citenamefont {Fazio}, \citenamefont {Soavi}, \citenamefont {Dutta}, \citenamefont {Balci}, \citenamefont {Shinde}, \citenamefont {Zhang}, \citenamefont {Katiyar}, \citenamefont {Keatley}, \citenamefont {Srivastava}, \citenamefont {Sankaranarayanan}, \citenamefont {Ferrari},\ and\ \citenamefont {Bhatia}}]{2021(dwivedi)graphene}%
  \BibitemOpen
  \bibfield  {author} {\bibinfo {author} {\bibfnamefont {N.}~\bibnamefont {Dwivedi}}, \bibinfo {author} {\bibfnamefont {A.~K.}\ \bibnamefont {Ott}}, \bibinfo {author} {\bibfnamefont {K.}~\bibnamefont {Sasikumar}}, \bibinfo {author} {\bibfnamefont {C.}~\bibnamefont {Dou}}, \bibinfo {author} {\bibfnamefont {R.~J.}\ \bibnamefont {Yeo}}, \bibinfo {author} {\bibfnamefont {B.}~\bibnamefont {Narayanan}}, \bibinfo {author} {\bibfnamefont {U.}~\bibnamefont {Sassi}}, \bibinfo {author} {\bibfnamefont {D.~D.}\ \bibnamefont {Fazio}}, \bibinfo {author} {\bibfnamefont {G.}~\bibnamefont {Soavi}}, \bibinfo {author} {\bibfnamefont {T.}~\bibnamefont {Dutta}}, \bibinfo {author} {\bibfnamefont {O.}~\bibnamefont {Balci}}, \bibinfo {author} {\bibfnamefont {S.}~\bibnamefont {Shinde}}, \bibinfo {author} {\bibfnamefont {J.}~\bibnamefont {Zhang}}, \bibinfo {author} {\bibfnamefont {A.~K.}\ \bibnamefont {Katiyar}}, \bibinfo {author} {\bibfnamefont {P.~S.}\ \bibnamefont {Keatley}}, \bibinfo {author} {\bibfnamefont {A.~K.}\ \bibnamefont {Srivastava}}, \bibinfo {author} {\bibfnamefont {S.~K. R.~S.}\ \bibnamefont {Sankaranarayanan}}, \bibinfo {author} {\bibfnamefont {A.~C.}\ \bibnamefont {Ferrari}},\ and\ \bibinfo {author} {\bibfnamefont {C.~S.}\ \bibnamefont {Bhatia}},\ }\bibfield  {title} {\bibinfo {title} {Graphene overcoats for ultra-high storage density magnetic media},\ }\href {https://doi.org/10.1038/s41467-021-22687-y} {\bibfield  {journal} {\bibinfo  {journal} {Nat. Commun.}\ }\textbf {\bibinfo {volume} {12}},\ \bibinfo {pages} {2854} (\bibinfo {year} {2021})}\BibitemShut {NoStop}%
\bibitem [{\citenamefont {Vansteenkiste}\ \emph {et~al.}(2014)\citenamefont {Vansteenkiste}, \citenamefont {Leliaert}, \citenamefont {Dvornik}, \citenamefont {Helsen}, \citenamefont {{Garcia-Sanchez}},\ and\ \citenamefont {Van~Waeyenberge}}]{2014(vansteenkiste)design}%
  \BibitemOpen
  \bibfield  {author} {\bibinfo {author} {\bibfnamefont {A.}~\bibnamefont {Vansteenkiste}}, \bibinfo {author} {\bibfnamefont {J.}~\bibnamefont {Leliaert}}, \bibinfo {author} {\bibfnamefont {M.}~\bibnamefont {Dvornik}}, \bibinfo {author} {\bibfnamefont {M.}~\bibnamefont {Helsen}}, \bibinfo {author} {\bibfnamefont {F.}~\bibnamefont {{Garcia-Sanchez}}},\ and\ \bibinfo {author} {\bibfnamefont {B.}~\bibnamefont {Van~Waeyenberge}},\ }\bibfield  {title} {\bibinfo {title} {The design and verification of {{MuMax3}}},\ }\href {https://doi.org/10.1063/1.4899186} {\bibfield  {journal} {\bibinfo  {journal} {AIP Adv.}\ }\textbf {\bibinfo {volume} {4}},\ \bibinfo {pages} {107133} (\bibinfo {year} {2014})}\BibitemShut {NoStop}%
\bibitem [{\citenamefont {Rannala}\ \emph {et~al.}(2022)\citenamefont {Rannala}, \citenamefont {Meo}, \citenamefont {Ruta}, \citenamefont {Pantasri}, \citenamefont {Chantrell}, \citenamefont {Chureemart},\ and\ \citenamefont {Chureemart}}]{2022(rannala)models}%
  \BibitemOpen
  \bibfield  {author} {\bibinfo {author} {\bibfnamefont {S.}~\bibnamefont {Rannala}}, \bibinfo {author} {\bibfnamefont {A.}~\bibnamefont {Meo}}, \bibinfo {author} {\bibfnamefont {S.}~\bibnamefont {Ruta}}, \bibinfo {author} {\bibfnamefont {W.}~\bibnamefont {Pantasri}}, \bibinfo {author} {\bibfnamefont {R.}~\bibnamefont {Chantrell}}, \bibinfo {author} {\bibfnamefont {P.}~\bibnamefont {Chureemart}},\ and\ \bibinfo {author} {\bibfnamefont {J.}~\bibnamefont {Chureemart}},\ }\bibfield  {title} {\bibinfo {title} {Models of advanced recording systems: {{A}} multi-timescale micromagnetic code for granular thin film magnetic recording systems},\ }\href {https://doi.org/10.1016/j.cpc.2022.108462} {\bibfield  {journal} {\bibinfo  {journal} {Comput. Phys. Commun.}\ }\textbf {\bibinfo {volume} {279}},\ \bibinfo {pages} {108462} (\bibinfo {year} {2022})}\BibitemShut {NoStop}%
\bibitem [{\citenamefont {Evans}\ \emph {et~al.}(2012{\natexlab{b}})\citenamefont {Evans}, \citenamefont {Hinzke}, \citenamefont {Atxitia}, \citenamefont {Nowak}, \citenamefont {Chantrell},\ and\ \citenamefont {{Chubykalo-Fesenko}}}]{2012(evans)stochastic}%
  \BibitemOpen
  \bibfield  {author} {\bibinfo {author} {\bibfnamefont {R.~F.~L.}\ \bibnamefont {Evans}}, \bibinfo {author} {\bibfnamefont {D.}~\bibnamefont {Hinzke}}, \bibinfo {author} {\bibfnamefont {U.}~\bibnamefont {Atxitia}}, \bibinfo {author} {\bibfnamefont {U.}~\bibnamefont {Nowak}}, \bibinfo {author} {\bibfnamefont {R.~W.}\ \bibnamefont {Chantrell}},\ and\ \bibinfo {author} {\bibfnamefont {O.}~\bibnamefont {{Chubykalo-Fesenko}}},\ }\bibfield  {title} {\bibinfo {title} {Stochastic form of the {{Landau-Lifshitz-Bloch}} equation},\ }\href {https://doi.org/10.1103/PhysRevB.85.014433} {\bibfield  {journal} {\bibinfo  {journal} {Phys. Rev. B}\ }\textbf {\bibinfo {volume} {85}},\ \bibinfo {pages} {014433} (\bibinfo {year} {2012}{\natexlab{b}})}\BibitemShut {NoStop}%
\bibitem [{\citenamefont {Ogawa}\ \emph {et~al.}(2024)\citenamefont {Ogawa}, \citenamefont {Bolyachkin}, \citenamefont {Dilipan}, \citenamefont {Kulesh}, \citenamefont {{Sepehri-Amin}},\ and\ \citenamefont {Takahashi}}]{2024(ogawa)exchangecoupled}%
  \BibitemOpen
  \bibfield  {author} {\bibinfo {author} {\bibfnamefont {D.}~\bibnamefont {Ogawa}}, \bibinfo {author} {\bibfnamefont {A.}~\bibnamefont {Bolyachkin}}, \bibinfo {author} {\bibfnamefont {A.~R.}\ \bibnamefont {Dilipan}}, \bibinfo {author} {\bibfnamefont {N.}~\bibnamefont {Kulesh}}, \bibinfo {author} {\bibfnamefont {H.}~\bibnamefont {{Sepehri-Amin}}},\ and\ \bibinfo {author} {\bibfnamefont {Y.~K.}\ \bibnamefont {Takahashi}},\ }\bibfield  {title} {\bibinfo {title} {Exchange-coupled {{Fe-Pt}}/{{Ru}}/{{Fe-Pt}} nanogranular films as potential heat-assisted-magnetic-recording media with reduced writing temperature},\ }\href {https://doi.org/10.1103/PhysRevApplied.22.054060} {\bibfield  {journal} {\bibinfo  {journal} {Phys. Rev. Applied}\ }\textbf {\bibinfo {volume} {22}},\ \bibinfo {pages} {054060} (\bibinfo {year} {2024})}\BibitemShut {NoStop}%
\bibitem [{\citenamefont {Okamoto}\ \emph {et~al.}(2002)\citenamefont {Okamoto}, \citenamefont {Kikuchi}, \citenamefont {Kitakami}, \citenamefont {Miyazaki}, \citenamefont {Shimada},\ and\ \citenamefont {Fukamichi}}]{2002(okamoto)chemicalorderdependent}%
  \BibitemOpen
  \bibfield  {author} {\bibinfo {author} {\bibfnamefont {S.}~\bibnamefont {Okamoto}}, \bibinfo {author} {\bibfnamefont {N.}~\bibnamefont {Kikuchi}}, \bibinfo {author} {\bibfnamefont {O.}~\bibnamefont {Kitakami}}, \bibinfo {author} {\bibfnamefont {T.}~\bibnamefont {Miyazaki}}, \bibinfo {author} {\bibfnamefont {Y.}~\bibnamefont {Shimada}},\ and\ \bibinfo {author} {\bibfnamefont {K.}~\bibnamefont {Fukamichi}},\ }\bibfield  {title} {\bibinfo {title} {Chemical-order-dependent magnetic anisotropy and exchange stiffness constant of {{FePt}} (001) epitaxial films},\ }\href {https://doi.org/10.1103/PhysRevB.66.024413} {\bibfield  {journal} {\bibinfo  {journal} {Phys. Rev. B}\ }\textbf {\bibinfo {volume} {66}},\ \bibinfo {pages} {024413} (\bibinfo {year} {2002})}\BibitemShut {NoStop}%
\bibitem [{\citenamefont {Rong}\ \emph {et~al.}(2006)\citenamefont {Rong}, \citenamefont {Li}, \citenamefont {Nandwana}, \citenamefont {Poudyal}, \citenamefont {Ding}, \citenamefont {Wang}, \citenamefont {Zeng},\ and\ \citenamefont {Liu}}]{2006(rong)sizedependent}%
  \BibitemOpen
  \bibfield  {author} {\bibinfo {author} {\bibfnamefont {C.-b.}\ \bibnamefont {Rong}}, \bibinfo {author} {\bibfnamefont {D.}~\bibnamefont {Li}}, \bibinfo {author} {\bibfnamefont {V.}~\bibnamefont {Nandwana}}, \bibinfo {author} {\bibfnamefont {N.}~\bibnamefont {Poudyal}}, \bibinfo {author} {\bibfnamefont {Y.}~\bibnamefont {Ding}}, \bibinfo {author} {\bibfnamefont {Z.~L.}\ \bibnamefont {Wang}}, \bibinfo {author} {\bibfnamefont {H.}~\bibnamefont {Zeng}},\ and\ \bibinfo {author} {\bibfnamefont {J.~P.}\ \bibnamefont {Liu}},\ }\bibfield  {title} {\bibinfo {title} {Size-{{Dependent Chemical}} and {{Magnetic Ordering}} in {{\textit{L}1\textsubscript{0}-FePt Nanoparticles}}},\ }\href {https://doi.org/10.1002/adma.200601904} {\bibfield  {journal} {\bibinfo  {journal} {Adv. Mater.}\ }\textbf {\bibinfo {volume} {18}},\ \bibinfo {pages} {2984} (\bibinfo {year} {2006})}\BibitemShut {NoStop}%
\bibitem [{\citenamefont {Hovorka}\ \emph {et~al.}(2012)\citenamefont {Hovorka}, \citenamefont {Devos}, \citenamefont {Coopman}, \citenamefont {Fan}, \citenamefont {Aas}, \citenamefont {Evans}, \citenamefont {Chen}, \citenamefont {Ju},\ and\ \citenamefont {Chantrell}}]{2012(hovorka)curie}%
  \BibitemOpen
  \bibfield  {author} {\bibinfo {author} {\bibfnamefont {O.}~\bibnamefont {Hovorka}}, \bibinfo {author} {\bibfnamefont {S.}~\bibnamefont {Devos}}, \bibinfo {author} {\bibfnamefont {Q.}~\bibnamefont {Coopman}}, \bibinfo {author} {\bibfnamefont {W.~J.}\ \bibnamefont {Fan}}, \bibinfo {author} {\bibfnamefont {C.~J.}\ \bibnamefont {Aas}}, \bibinfo {author} {\bibfnamefont {R.~F.~L.}\ \bibnamefont {Evans}}, \bibinfo {author} {\bibfnamefont {X.}~\bibnamefont {Chen}}, \bibinfo {author} {\bibfnamefont {G.}~\bibnamefont {Ju}},\ and\ \bibinfo {author} {\bibfnamefont {R.~W.}\ \bibnamefont {Chantrell}},\ }\bibfield  {title} {\bibinfo {title} {The {{Curie}} temperature distribution of {{FePt}} granular magnetic recording media},\ }\href {https://doi.org/10.1063/1.4740075} {\bibfield  {journal} {\bibinfo  {journal} {Appl. Phys. Lett.}\ }\textbf {\bibinfo {volume} {101}},\ \bibinfo {pages} {052406} (\bibinfo {year} {2012})}\BibitemShut {NoStop}%
\bibitem [{\citenamefont {Liu}\ \emph {et~al.}(2017)\citenamefont {Liu}, \citenamefont {Huang}, \citenamefont {Ju},\ and\ \citenamefont {Victora}}]{2017(liu)thermal}%
  \BibitemOpen
  \bibfield  {author} {\bibinfo {author} {\bibfnamefont {Z.}~\bibnamefont {Liu}}, \bibinfo {author} {\bibfnamefont {P.-W.}\ \bibnamefont {Huang}}, \bibinfo {author} {\bibfnamefont {G.}~\bibnamefont {Ju}},\ and\ \bibinfo {author} {\bibfnamefont {R.~H.}\ \bibnamefont {Victora}},\ }\bibfield  {title} {\bibinfo {title} {Thermal switching probability distribution of {{L10 FePt}} for heat assisted magnetic recording},\ }\href {https://doi.org/10.1063/1.4983033} {\bibfield  {journal} {\bibinfo  {journal} {Appl. Phys. Lett.}\ }\textbf {\bibinfo {volume} {110}},\ \bibinfo {pages} {182405} (\bibinfo {year} {2017})}\BibitemShut {NoStop}%
\bibitem [{\citenamefont {Richter}\ and\ \citenamefont {Parker}(2017)}]{2017(richter)temperature}%
  \BibitemOpen
  \bibfield  {author} {\bibinfo {author} {\bibfnamefont {H.~J.}\ \bibnamefont {Richter}}\ and\ \bibinfo {author} {\bibfnamefont {G.~J.}\ \bibnamefont {Parker}},\ }\bibfield  {title} {\bibinfo {title} {Temperature dependence of the anisotropy field of {{L1\textsubscript{0} FePt}} near the {{Curie}} temperature},\ }\href {https://doi.org/10.1063/1.4984911} {\bibfield  {journal} {\bibinfo  {journal} {J. Appl. Phys.}\ }\textbf {\bibinfo {volume} {121}},\ \bibinfo {pages} {213902} (\bibinfo {year} {2017})}\BibitemShut {NoStop}%
\bibitem [{\citenamefont {Callen}\ and\ \citenamefont {Callen}(1966)}]{1966(callen)present}%
  \BibitemOpen
  \bibfield  {author} {\bibinfo {author} {\bibfnamefont {H.}~\bibnamefont {Callen}}\ and\ \bibinfo {author} {\bibfnamefont {E.}~\bibnamefont {Callen}},\ }\bibfield  {title} {\bibinfo {title} {The {{Present Status}} of the {{Temperature Dependence}} of {{Magnetocrystalline Anisotropy}}, and the \textit{l}(\textit{l}+1)2 {{Power Law}}},\ }\href {https://doi.org/10.1016/0022-3697(66)90012-6} {\bibfield  {journal} {\bibinfo  {journal} {J. Phys. Chem. Solids}\ }\textbf {\bibinfo {volume} {27}},\ \bibinfo {pages} {1271} (\bibinfo {year} {1966})}\BibitemShut {NoStop}%
\bibitem [{\citenamefont {Mryasov}\ \emph {et~al.}(2005)\citenamefont {Mryasov}, \citenamefont {Nowak}, \citenamefont {Guslienko},\ and\ \citenamefont {Chantrell}}]{2005(mryasov)temperaturedependent}%
  \BibitemOpen
  \bibfield  {author} {\bibinfo {author} {\bibfnamefont {O.~N.}\ \bibnamefont {Mryasov}}, \bibinfo {author} {\bibfnamefont {U.}~\bibnamefont {Nowak}}, \bibinfo {author} {\bibfnamefont {K.~Y.}\ \bibnamefont {Guslienko}},\ and\ \bibinfo {author} {\bibfnamefont {R.~W.}\ \bibnamefont {Chantrell}},\ }\bibfield  {title} {\bibinfo {title} {Temperature-dependent magnetic properties of {{FePt}}: {{Effective}} spin {{Hamiltonian}} model},\ }\href {https://doi.org/10.1209/epl/i2004-10404-2} {\bibfield  {journal} {\bibinfo  {journal} {Europhys. Lett.}\ }\textbf {\bibinfo {volume} {69}},\ \bibinfo {pages} {805} (\bibinfo {year} {2005})}\BibitemShut {NoStop}%
\bibitem [{\citenamefont {Zhu}\ and\ \citenamefont {Li}(2013)}]{2013(zhu)understanding}%
  \BibitemOpen
  \bibfield  {author} {\bibinfo {author} {\bibfnamefont {J.-G.}\ \bibnamefont {Zhu}}\ and\ \bibinfo {author} {\bibfnamefont {H.}~\bibnamefont {Li}},\ }\bibfield  {title} {\bibinfo {title} {Understanding {{Signal}} and {{Noise}} in {{Heat Assisted Magnetic Recording}}},\ }\href {https://doi.org/10.1109/TMAG.2012.2231855} {\bibfield  {journal} {\bibinfo  {journal} {IEEE Trans. Magn.}\ }\textbf {\bibinfo {volume} {49}},\ \bibinfo {pages} {765} (\bibinfo {year} {2013})}\BibitemShut {NoStop}%
\bibitem [{\citenamefont {Everaert}\ \emph {et~al.}(2023)\citenamefont {Everaert}, \citenamefont {Van~Waeyenberge}, \citenamefont {Wiekhorst},\ and\ \citenamefont {Leliaert}}]{2023(everaert)impact}%
  \BibitemOpen
  \bibfield  {author} {\bibinfo {author} {\bibfnamefont {K.}~\bibnamefont {Everaert}}, \bibinfo {author} {\bibfnamefont {B.}~\bibnamefont {Van~Waeyenberge}}, \bibinfo {author} {\bibfnamefont {F.}~\bibnamefont {Wiekhorst}},\ and\ \bibinfo {author} {\bibfnamefont {J.}~\bibnamefont {Leliaert}},\ }\bibfield  {title} {\bibinfo {title} {The impact of temperature on thermal fluctuations in magnetic nanoparticle systems},\ }\href {https://doi.org/10.1063/5.0147434} {\bibfield  {journal} {\bibinfo  {journal} {Appl. Phys. Lett.}\ }\textbf {\bibinfo {volume} {122}},\ \bibinfo {pages} {211902} (\bibinfo {year} {2023})}\BibitemShut {NoStop}%
\bibitem [{\citenamefont {Vogler}\ \emph {et~al.}(2016)\citenamefont {Vogler}, \citenamefont {Abert}, \citenamefont {Bruckner}, \citenamefont {Suess},\ and\ \citenamefont {Praetorius}}]{2016(vogler)basic}%
  \BibitemOpen
  \bibfield  {author} {\bibinfo {author} {\bibfnamefont {C.}~\bibnamefont {Vogler}}, \bibinfo {author} {\bibfnamefont {C.}~\bibnamefont {Abert}}, \bibinfo {author} {\bibfnamefont {F.}~\bibnamefont {Bruckner}}, \bibinfo {author} {\bibfnamefont {D.}~\bibnamefont {Suess}},\ and\ \bibinfo {author} {\bibfnamefont {D.}~\bibnamefont {Praetorius}},\ }\bibfield  {title} {\bibinfo {title} {Basic noise mechanisms of heat-assisted-magnetic recording},\ }\href {https://doi.org/10.1063/1.4964949} {\bibfield  {journal} {\bibinfo  {journal} {J. Appl. Phys.}\ }\textbf {\bibinfo {volume} {120}},\ \bibinfo {pages} {153901} (\bibinfo {year} {2016})}\BibitemShut {NoStop}%
\end{thebibliography}%

\end{document}